\documentclass[twocolumn,superscriptaddress,nofootinbib,amsmath,amssymb,amsfonts]{aastex631}
\usepackage{newtxtext,newtxmath}
\usepackage[T1]{fontenc}
\usepackage{ae,aecompl}
\usepackage{multirow}
\usepackage{rotating}

\usepackage{graphicx}	
\usepackage{amsmath}
\usepackage{url}
\usepackage{color}
\usepackage{indentfirst}

\usepackage[normalem]{ulem}

\newcommand{\new}[1]{#1}
\newcommand{\eg}{e.g.,~}
\newcommand{\ie}{i.e.,~}
\usepackage{hyperref}
\hypersetup{
  colorlinks=true,        % false: boxed links; true: colored links
  linkcolor=blue,         % color of internal links
  citecolor=cyan,         % color of links to bibliography
  urlcolor=cyan           % color of external links
}

\begin{document}

\title{``Extended emission'' from fallback
  accretion onto merger remnants}

\author[0000-0002-9955-3451]{Carlo Musolino}
\affiliation{Institut f\"ur Theoretische Physik, Goethe Universit\"at, Max-von-Laue-Straße 1, 60438 Frankfurt am Main, Germany}
\author[0000-0002-9062-3775]{Rapha\"el Duqu\'e}
\affiliation{Institut f\"ur Theoretische Physik, Goethe Universit\"at, Max-von-Laue-Straße 1, 60438 Frankfurt am Main, Germany}
\author[0000-0002-1330-7103]{Luciano Rezzolla}
\affiliation{Institut f\"ur Theoretische Physik, Goethe Universit\"at, Max-von-Laue-Straße 1, 60438 Frankfurt am Main, Germany}
\affiliation{School of Mathematics, Trinity College, Dublin 2, Ireland}
\affiliation{Frankfurt Institute for Advanced Studies, Ruth-Moufang-Str. 1, 60438 Frankfurt am Main, Germany}

\date{\today}

\begin{abstract}
  Using a set of general-relativistic magnetohydrodynamics simulations
  that include proper neutrino transfer, we assess for the first time the
  role played by the fallback accretion onto the remnant from a binary
  neutron-star merger over a timescale of hundreds of seconds. In
  particular, we find that, independently of the equation of state, the
  properties of the binary, and the fate of the remnant, the fallback
  material reaches a total mass of $\gtrsim 10^{-3}\,M_\odot$, \ie about
  50\% of the unbound matter, and that the fallback accretion rate
  follows a power-law in time with slope $\sim t^{-5/3}$. Interestingly,
  the timescale of the fallback and the corresponding accretion
  luminosity are in good agreement with the so-called ``extended
  emission'' observed in short gamma-ray bursts (GRBs). Using a simple
  electromagnetic emission model based on the self-consistent
  thermodynamical state of the fallback material heated by r-process
  nucleosynthesis, we show that this fallback material can shine in the
  gamma- and X-rays with luminosities $\gtrsim \,10^{48}\,{\rm erg/s}$
  for hundreds of seconds, thus making it a good and natural candidate
  to explain the extended emission in short GRBs. In addition, our model
  for the emission by the fallback material reproduces well and rather
  naturally some of the phenomenological traits of the extended emission,
  such as its softer spectra with respect to the prompt emission and the
  presence of exponential cutoffs in time. Our results clearly highlight
  that fallback flows onto merger remnants cannot be neglected and the
  corresponding emission represents a very promising and largely
  unexplored avenue to explain the complex phenomenology of GRBs.
\end{abstract}


%\maketitle

%-------------------------------------------------------------------
%-------------------------------------------------------------------
\section{Introduction}
\label{sec:intro}
%-------------------------------------------------------------------
%-------------------------------------------------------------------

The merger of binary compact objects is a complex phenomenon involving
many different physical processes, making them a great laboratory for
many branches of physics and astrophysics~\citep{Baiotti2016,
  Paschalidis2016}. Nonetheless, because the inspiral and merger
processes are deterministic, any observable produced by this process has
the potential of revealing the properties of the progenitor binary and,
hence, of matter and gravity under extreme conditions. In particular,
while the characteristics of the different outflows launched in the
merger and their electromagnetic (EM) signals are fully determined by the
properties of the binary, \new{the self-consistent and long-term
  modelling of these events and their observational counterparts is
  extremely complex and currently cannot be performed without a number of
  approximations}. The first binary neutron-star (BNS) multi-messenger
merger event, GW170817, presented, in addition to gravitational waves,
also a rich set of EM counterparts: a weak short gamma-ray burst (GRB), a
thermal kilonova transient and a long-lasting afterglow associated with
the relativistic jet launched by the remnant~\citep{Abbott2017_etal,
  Abbott2017b, Abbott2017dddd} . This afterglow could be currently
re-brightening in the X-rays and changing spectral index, possibly
revealing another emission component~\citep{Balasubramanian2021,
  Hajela2022_ag}. Other EM counterparts are predicted from BNS mergers,
such as early X-ray flash,~\citep{Fernandez2016_em}, but have not yet
been observed. These EM signals from BNS mergers are part of the rich
phenomenology of short GRBs that has been explored well before GW170817
\citep[see, \eg][]{Berger2013b,DAvanzo2015}. In this phenomenology, and
the so-called ``extended emission'' episodes~\citep{Norris2006_ee} --
that is, stretches of \new{soft} gamma-ray \new{to hard X-ray} emission
that can last for hundreds of seconds after the prompt emission of short
GRBs -- a particularly puzzling aspect and are the main focus of this
paper. \new{It is important to note that the extended-emission component
  is clearly distinct from the prompt emission because of its softer
  spectrum, as well as from the afterglow.} In their ensemble, all of
these counterparts likely arise from distinct outflows from the merger,
either at ultrarelativistic or at more moderate speeds -- and the
relative importance of these outflows is therefore determined by the
properties of the progenitor binary.

In this rich but complex picture, the ideal tool to study the full
sequence from merger to EM signals is first-principle
general-relativistic magnetohydrodynamical (GRMHD) simulations of the
merger followed by an adequate post-processing to implement the emission
models of the EM counterparts. Studying the full picture from the binary
properties to the EM signals has many advantages. First, using
constraints on the binary system provided by low-latency analysis of the
gravitational-wave data (such as the chirp mass) can guide follow-up
searches of EM counterparts by partially predicting the expected
signals. This has already been done for the kilonova
signal~\citep{Barbieri2020_kn, Nicholl2021_kn}, and should be extended to
other counterparts, most notably, those detected in far-away or on-axis
events expected for the future, such as the afterglow~\citep[see,
  \eg][]{Duque2019_radio}. An early prediction of EM counterparts would
obviously be useful to optimise the follow-up searches and, in the
third-generation detectors era~\citep{Punturo:2010}, in which many BNSs
will be detected per day~\citep[see, \eg][]{Ronchini2022_et}, to choose
the most promising systems EM-wise to focus the follow-up efforts on
those. Second, studying existing catalogues of EM signals from BNS
mergers, such as short GRBs and their afterglows can help to constrain
the progenitors of short GRBs and their formation channels. This
approach, together with population studies, can constrain, for instance,
the GRB-luminosity function~\citep{Ghirlanda2016_pop} by linking it to
the progenitor system parameters. Finally, ab-initio calculations
represent possibly the only way in which it may be possible to find
conclusive answer to some of the most puzzling aspects of the EM
phenomenology of short GRBs, such as the extended emission.
\new{Moreover, gravitational-wave triggered events may enable us to probe
  the off-axis properties of the extended emission, which are mostly
  inaccessible through sGRB catalogs, but might be very useful in
  constraining the emission associated with these events.}

We recall that the extended emission in the gamma-ray and X-ray bands
covers two recurring features in short GRBs observed. As already
mentioned, the gamma-ray extended emission is characterized by an episode
of gamma-ray emission that can last for hundreds of seconds after the
prompt emission of short GRBs and that is of lower luminosity and softer
spectrum than the parent prompt (this is also referred to as the
``spike''). Gamma-ray extended emission was first discovered in the
prototypical GRBs\,050724 and 060614~\citep{Barthelmy2005,
  Gehrels2006_ee}, that were initially classified as long GRBs because of
their extended emission. Gamma-ray extended emission is present in
$2-25\%$ of short GRBs, depending on the instrument and search method,
and bursts with extended emission are characterized by very low spectral
lags in the parent spike~\citep{Norris2006_ee, Kaneko2015_ee},
\new{consistent with those of sGRBs}. Short GRBs with and without
detectable extended emission are indistinguishable in terms of galactic
offset and host galaxy properties~\citep{Fong2022, Nugent2022},
consistent with them arising from the same progenitors systems, perhaps
with particular characteristics.

On the other hand, the extended emission in the X-ray band was discovered
in a sample of \textit{Swift} short GRBs with early X-ray observations
and extended gamma-rays observations~\citep{Kagawa2015_xray}. These
authors concluded from a restricted sample that the extended near-flat
X-ray emission was the X-ray counterpart of the extended emission in the
gamma-rays. The X-ray extended emission has a similar duration (up to
hundreds of seconds) and, within the sample studied, featured an
exponential cutoff. As a possible and simple mirror of the extended
emission in gamma-rays, X-ray extended emission was studied
\textit{per-se}, and found in approximately 20-50\% of short GRBs
detected by \textit{Swift}~\citep{Kisaka2017,Kagawa2019_ee}. In
short GRBs with measured redshift, the luminosities of extended emission
episodes in the X-rays spans from $10^{46}$ to $10^{49}$\,erg/s in the
\textit{Swift}/XRT band~\citep{Kisaka2017}. It should be remarked
that the fraction of the total energy lost in a short GRB with extended
emission that is output in the extended-emission episode varies
widely~\citep{Perley2009_ee, Bostanci2013_ee, Kagawa2019_ee}, and the
extended-emission episode itself can largely exceed the fluence of the
prompt emission, suggesting a large reservoir of energy active on long
timescales.

Extended emission in short GRBs has been interpreted through various
pictures. \new{The list of proposed models today includes} dissipation in
a second jet launched by the black-hole central engine through disk
accretion~\citep{Barkov2011_ee, Gottlieb:2023b} or through material
falling back after the burst~\citep{Kisaka2015_jets}, or from a magnetar
central engine~\citep{Bucciantini2012_ee}. Magnetar models also consider
the relativistic wind as a potential source \citep{Metzger2008,
  Murase2018}, or the interaction of the slow and fast winds launched
respectively before and after the dissipation of the differential
rotation of the merger remnant~\citep{Rezzolla2014b,
  Ciolfi2014}. Scenarios invoked for longer-duration soft episodes of
long GRBs or of flares in GRB afterglows can likely be adapted to
extended emission in short GRBs \citep[see,
  \eg][]{Lee2009_ee,Gao2022_211211A}

The ejection of material that remains gravitationally \textit{bound} to
the central object and falls back on long time-scales seems to be
ubiquitous in explosive events. Such material is invoked in various
contexts, from tidal disruption events (TDEs) with a mostly tidal
component~\citep{Krolik2012}, to BNS mergers considering secular
ejecta~\citep{Ishizaki2021_fba}. The dynamics of such material has
attracted much attention, starting from the early analytical studies
prescribing the distribution of energy in the debris~\citep{Rees1988},
and notoriously leading to a simple mass-accretion rates scaling as $\sim
t^{-5/3}$ with time, to numerical-relativity simulations that have
refined this picture under more realistic conditions~\citep[see,
  \eg][]{Rosswog2007_fb}. TDEs, in particular, have a rich observational
history, with X-ray signatures following the $\sim t^{-5/3}$ accretion
rate to more complex phenomenology hinting to the launching of
jets~\citep[see, \eg][]{Komossa2015}.

For BNS mergers, the picture for fallback material is somewhat
complicated by the presence of various ejecta (that are either dynamical
or secular~\citep{Gill2019}), by the nature of the remnant compact object
(which can either be a massive neutron star or a black
hole)~\citep{Nathanail2021}, and by the interaction of the jet with the
ejected material~\citep{Murguia-Berthier2016, Urrutia2020, Nathanail2020c, 
Gottlieb2022, Hamidani2020, Kiuchi2023b, Mpisketzis2024}.
Recently,~\citet{Metzger2021_fb} and \citet{Ishizaki2021_fba} have sought
to explain the putative excess in the afterglow of GW170817 with the
emission of fallback material from the secular ejecta.
Furthermore,~\citet{Ishizaki2021_fbb} have studied the impact of nuclear
reaction reigniting in the fallback accretion, showing a halt in the
accretion flow due to the internal heating. However, while interesting,
these studies lack a first-principles insight into the amount of bound
ejecta and on the fallback dynamics starting from the merger phase, as
well as a study of the emission from this material in the shorter
timescales relevant for extended emission.

\begin{table*}
\center
  \begin{tabular}{l|c|c|c|c|c|c|c|c|c|c|c}
  \hline
                       & $M_1$      & $M_2$      & $q$ & EOS     & $t_{\rm coll}$ & remnant   & $M_{\rm rem}$ & $m_{\rm tot}$      & $m_{\rm FB}$ & $m_{\rm KN}$  & $m_{\rm FB} / m_{\rm KN}$\\ 
                       & [$M_\odot$] & [$M_\odot$] &      &        & [${\rm ms}$] &           & [$M_\odot$]  & [$M_\odot$]         & [$M_\odot$]  & [$M_\odot$]   & [$\%$]\\ \hline
  \texttt{SFHO-q1.0}   & 1.36       & 1.36       & 1.00 & SFHO   & 3.7          & black hole & 2.63        & $8.62 \times 10^{-3}$ & $1.60 \times 10^{-3}$ & $7.02 \times 10^{-3}$ & 23\\
  \texttt{SFHO-q.75}   & 1.18       & 1.57       & 0.75 & SFHO   & 4.0          & black hole & 2.64        & $1.04 \times 10^{-2}$ & $3.48 \times 10^{-3}$ & $6.92 \times 10^{-3}$ & 50\\
  \texttt{TNTYST-q1.0} & 1.36       & 1.36       & 1.00 & TNTYST & --           & HMNS       & 2.68        & $9.97 \times 10^{-3}$ & $3.76 \times 10^{-3}$ & $6.21 \times 10^{-3}$ & 60\\
  \texttt{TNTYST-q.75} & 1.18       & 1.57       & 0.75 & TNTYST & --           & HMNS       & 2.78        & $1.33 \times 10^{-2}$ & $4.05 \times 10^{-3}$ & $9.25 \times 10^{-3}$ & 44\\
  \hline
\end{tabular}
\caption{The four binary neutron-star systems considered in this study,
  with binary parameters and general characteristics of the fallback
  episodes. $M_1, M_2$: masses of the component neutron stars in the
  binary. $q = M_1 / M_2 < 1$: mass ratio of the binary. $t_{\rm coll}$:
  in the case that a black-hole forms in the simulation domain, time of
  formation after merger. $M_{\rm rem}$: mass of the remnant compact
  object of the merger. $m_{\rm tot}$: total mass ejected from pre-merger
  to post-merger phases. $m_{\rm FB}$: mass of bound material
  ejected. $m_{\rm KN}$: mass of unbound material ejected, thus denoted
  for its role in the kilonova signal.}
\label{tab:1}
\end{table*}

One of the main scopes of this paper is to improve on these studies by
using state-of-the GRMHD simulations with proper neutrino transport
followed by a semi-analytical treatment of the fallback dynamics to study
in detail the fallback accretion and the radiation arising from this
inflow. This allows us to explore the possibility that extended emission
can be attributed to this fallback material. We contrast the properties
of this emission component according to the mass-ratio and equation of
state (EOS) of the material assumed in the neutron stars, in a first
search for phenomenological trends with binary characteristics. In this
way, we find that the amount of bound matter is \textit{substantial},
being almost $50\%$ of the dynamically unbound matter and reaching a total of
$\gtrsim 10^{-3}\,M_\odot$. Furthermore, the accretion rate follows a
universal power-law in time with slope $\simeq t^{-5/3}$, which is
independent of the EOS, the properties of the binary and the fate of the
remnant. Importantly, the timescale of the fallback and the corresponding
accretion luminosity all are in good agreement with the long-term
emission observed in short GRBs. Using a simple EM emission model based
on the thermodynamical state of the fallback material heated by r-process
nucleosynthesis, we show that this fallback material can shine in the
gamma- and X-rays with luminosities $\gtrsim \,10^{48}\,{\rm erg/s}$ for
hundreds of seconds, thus making it as a good candidate to explain the
extended emission and reproducing rather naturally some of the
phenomenological traits of extended-emission, such as its softer spectra
with respect to the prompt emission and the presence of exponential
cutoffs in time.

The plan of the paper is as follows. Sec. \ref{sec:2} presents the
framework of our numerical simulations, including the choice of the
binary systems we simulate, as well as a description of the material
ejected from the mergers we simulate. Sec. \ref{sec:3} presents our
semi-analytical model to predict the dynamics of the fallback of the
bound material ejected and the radiation emitted by this fallback flow.
In Sec. \ref{sec:discussion} we present our finding of this fallback
radiation as a new high-energy emission component of BNS mergers, and we
discuss this component in relation to the observations of extended
emission episodes of short GRBs. In Sec. \ref{sec:conclusion} we conclude
with an outlook of the role of this fallback radiation in other
long-lasting high-energy signals such as X-ray plateaus.

%-------------------------------------------------------------------
%-------------------------------------------------------------------
\section{Binary mergers and fallback material}
\label{sec:2}
%-------------------------------------------------------------------
%-------------------------------------------------------------------

%===================================================================
\subsection{Numerical Setup}
\label{sec:method}
%===================================================================

As mentioned in the Introduction, what distinguishes our work from
similar studies in the literature is the realistic modelling of binary
neutron-star mergers so as to have not only an accurate description of
the matter ejection on the nonlinear physics involved in the merger, but
also an accurate dependence of the fallback dynamics on the properties of
the binaries. To this scope, we have considered four representative
binaries described by two different and temperature-dependent EOSs. The
first one is the SFHO EOS \citep{Hempel2010}, which is based on
relativistic mean field calculations which predicts a maximum nonrotating
(TOV) mass of $M_{\rm TOV}\simeq 2.06\,M_{\odot}$; the second EOS is the
TNTYST~\citep{Togashi2017}, that is based on variational many-body
theory, and which instead allows for higher masses, with $M_{\rm TOV}
\simeq 2.23\,M_\odot$. For all of the four binaries, the total
gravitational mass at infinite separation exceeds the expected maximum
mass for a uniformly rotating remnant with their respective EOS assuming
a ratio $M_{\rm max}/M_{\rm TOV} \simeq1.2-1.25$~\citep{Breu2016,
  Musolino2023b}. On the other hand, all the binaries have the same chirp
mass corresponding to the estimated value for the GW170917 event
\citep{LIGOScientific:2017vwq} and considering $M_2~(M_1)$ to be the
gravitational mass of the secondary (primary) star in the binary, we
consider two different mass ratios $q:= M_2/M_1 \leq 1$, namely, $q=1.00$
and $q=0.75$ to model either an equal or an unequal-mas merger. Overall,
this choice of EOSs and mass ratios allows us to explore and contrast the
scenarios modelled by the simulation in which the merger remnant remains
a hypermassive neutron star (HMNS) from the one in which it collapses to
a black hole a few millisecond after the merger. Details on the
properties of the four binaries considered and of their outcome can be
found in Tab.~\ref{tab:1}.

As mentioned earlier, the evolution of the four binaries is carried out
making use of the state-of-the-art numerical 3+1 code-suite developed in
Frankfurt, which consists of the \texttt{FIL} code for the higher-order
finite-difference solution of the GRMHD equations using high-resolution
shock-capturing (HRSC) methods~\citep{Toro99, Rezzolla_book:2013} and a
fourth-order accurate conservative finite-difference
scheme~\citep{DelZanna2007}. The evolution of the spacetime is instead
carried out with the \texttt{Antelope} spacetime
solver~\citep{Most2019b}, which solves the constraint damping formulation
of the Z4 formulation of the Einstein equations~\citep{Bernuzzi:2009ex,
  Alic:2011a}. The code-suite evolves the full set of equations in
conjunction with the \texttt{EinsteinToolkit}~\citep{loeffler_2011_et,
  EinsteinToolkit_etal:2022_05}, exploiting the \texttt{Carpet}
box-in-box AMR driver in Cartesian coordinates~\citep{Schnetter:2006pg},

In addition, to accurately capture the temperature and composition of the
ejected and bound material, the effect of weak interactions needs to be
properly included. We do so by employing the recently developed code
\texttt{FIL-M1}~\citep{Musolino2023} which employs a moment-based
neutrino-transport scheme. More specifically, \texttt{FIL-M1} evolves the
first two moments of the energy-integrated Boltzmann equations for
neutrinos using standard, second-order finite-volume high-resolution
shock-capturing methods. Thanks to the moment-based radiation transport
scheme, we are able to include all the leading order weak reactions in
our simulations, thereby \new{accounting for neutrino heating and cooling and
  momentum transfer, as well as the composition changes resulting from
  weak interactions.}

All our simulations are performed on a grid spanning $\Omega=[-1500,
  1500]^3\,{\rm km}$ with seven fixed levels of refinement and the grid
spacing on the finest level being $dx=310\,{\rm m}$. Moreover, we include
the effect of magnetic fields in all our simulations initialising it as a
purely poloidal magnetic field confined inside the stars, with a maximum
initial strength of \new{$\simeq 10^{16.5}\,{\rm G}$ for the equal-mass
  systems \texttt{SFHO-q1.0} and \texttt{TNTYST-q1.0}, $\simeq
  10^{16.7}\,{\rm G}$ for \texttt{SFHO-q.75} and $\simeq 10^{15}\,{\rm
    G}$ for \texttt{TNTYST-q.75}}~\citep[see][for a recent discussion of
  why magnetic fields confined to the crust are not
  likely]{Chabanov2022}. This value, despite being much larger than what
would be expected for old neutron stars at the end of their inspiral, has
a corresponding magnetic energy that is much smaller than the binding
energy of the system and hence does not significantly affect the dynamics
of the system before merger~\citep[see, \eg][]{Giacomazzo:2009mp}. In
addition, and as customary in this type of simulations, starting with an
unrealistically large magnetic field allows us to better capture
realistic magnetic field values after the BNS merger despite our
resolution being insufficient for correctly capturing the dynamical
magnetic field amplification due to the Kelvin-Helmholtz instability.

%===================================================================
\subsection{Bound and unbound material}
\label{sec:baum}
%===================================================================

Clearly, a very important aspect of our research consists of properly
distinguishing matter that is gravitationally bound from matter that is
not and hence will not fallback onto the merger remnant. To this scope,
and as customary in these simulations, we construct a matter ``detector''
consisting of a spherical surface a distance $R_d = 300\,M_\odot$ from
the centre of coordinates and record the properties of the matter flowing
out of such a 2-sphere. In particular, we use the covariant time
component of the material's four-velocity $u_t$ to determine the matter
that is gravitationally bound (\ie $u_t > -1$) or unbound (\ie $u_t <
-1$). We favour this criterion, which is also referred to as the
``geodesic criterion'', over the other common alternative represented by
the so-called ``Bernoulli criterion'' $h u_t >
-1$~\citep{Rezzolla_book:2013}, where $h$ is the specific enthalpy,
because the pure kinematic quantity $u_t$ relates directly to the
subsequent dynamics of the bound material in the gravitational field of
the remnant, that we will use to determine the fallback time below.  We
note that, in principle, hydrodynamical processes can convert internal
energy to kinetic energy, thus possibly unbinding material that was
initial deemed bound through the $u_t$ criterion; similarly, matter that
is considered unbound can undergo shock heating and become bound
\citep[see][for a discussion]{Bovard2017}. However, as we will show in
Sec.~\ref{sec:slope} by varying the position of the matter detector, the
conversion of internal energy into kinetic energy does not occur in
practice in our simulations, likely because of to the small rest-mass
density of the matter once it reaches the detector.

Because the dynamics of the bound matter outside of the detection
sphere cannot be followed in detail (even if bound, the matter past
the detector at $R_d$ travels outwards to distances that are orders of
magnitude larger than $R_d$), we model it in terms of geodesic motion
in the gravitational field of the merger remnant. This is a reasonable
assumption given that the rest-mass densities past the detector are
comparatively very small (\ie $\rho \lesssim 10^{9}\,{\rm g/cm}^3$)
and hydrodynamical effects can be neglected. More importantly, and as
we show in Sec.~\ref{sec:eos}, the kinetic energy of the material
detected decreases over the course of the post-merger evolution for
all the systems studied. Thus, the material ejected at later times
does not have the opportunity to collide with matter ejected earlier
(potentially leading to shocks), so that we do not expect any
interaction between the trajectories of the bound material.

A most important property of the bound matter is the time it will take to
fallback onto the remnant. We could calculate this ``fallback timescale''
$t_{\rm FB}$ in terms of the geodesic equations for the matter collected
at the detector (and indeed we have done so) but a much simpler and
computationally less expensive estimate can be obtained using simple and
analytic Newtonian expressions. In particular, matter with 4-velocity
$u_t$ and specific angular momentum $\ell$, will have an orbit in the
remnant's gravitational field with mass $M_{\rm rem}$ with specific
energy
\begin{equation}
\epsilon := -(1 + u_t) < 0\,,
\label{eq:epsilon}
\end{equation}
orbital eccentricity $e$
\begin{equation}
e := \sqrt{1 + \frac{2 \epsilon \ell^2}{M_{\rm rem}^2}}\,,
\label{eq:ecc}
\end{equation}
and orbit semi-major axis $a$
\begin{equation}
a := -\frac{M_{\rm rem}}{2\epsilon}\,.
\end{equation}

In general, in its motion from the detector surface away from the merger
remnant and back, the ejected matter does not trace a full orbit and
duration of its free-fall orbit is given by
\begin{equation}
t_{\rm FB} := \frac{a ^2 (1 - e ^ 2)^2}{\ell} \int_{\theta_s}^{2\pi - \theta_s}
\frac{1}{ (1 + e  \cos \theta)^2} d \theta \,,
\label{eq:tfb}
\end{equation}
where the $\theta$ is the angle traced by the (planar) orbit and
$\theta_s$ can be determined from $a$, $e$ and $R_d$ \citep[see,
  \eg][]{Rosswog2007_fb}. This is the explicit solution to the Newtonian
equation of motion in the gravity field of the central object.

However, we have the particular aim here of describing fallback flows on
timescales of tens of seconds, much larger than the dynamical timescale
around the central object.  In this case, the duration of the fallback
orbit is almost that of the full Newtonian orbit, whose period is 
\begin{equation}
t_{\rm FB} = \frac{\pi}{\sqrt{2}}M_{\rm rem} |\epsilon|^{-3/2} \,.
\label{eq:orb}
\end{equation}
which is the well-know result of Newton's two-body problem.

Equation~\eqref{eq:tfb} (and~\eqref{eq:orb}) ultimately represents an
approximation of the actual fallback time that would be computed for the
geodesic orbit a massive test-particle having the same energy and angular
momentum moving in a Kerr spacetime with mass $M_{\rm rem}$ and
dimensionless spin $\chi_{\rm rem}$. In Appendix~\ref{sec:A}, we report a
comparison between the fallback times computed with the two
approaches. However, we can anticipate here that for long-period orbits,
\ie orbits with $t_{\rm FB} \gtrsim 1\,{\rm s}$, the differences between
the Kerr-geodetic estimate and the Newtonian estimate are always
$\lesssim 0.2\,\%$, which is not surprising given that the orbits are
mostly in a weak-field region where $r / M_{\rm rem} \gtrsim 200$. Such
an error is well below other uncertainties in our modelling
%% (see Sec.~\ref{sec:uncertainty} for a discussion)
and given that Eq.~\eqref{eq:tfb} comes with a considerable computational
gain, it has been set as the method of choice in our analysis. Note that,
by construction, we cannot monitor bound material with orbits that are
totally contained within the detector 2-sphere. Stated differently, the
fallback times we compute are necessarily larger than
\begin{equation}
  t_{\rm FB, min} \gtrsim 0.1\, \sqrt{
    \left(\frac{R_d}{300\,M_{\odot}}\right)^3
    \left(\frac{2.7\,M_{\odot}}{M_{\rm rem}}\right)
    } \quad {\rm s}\,,
\label{eq:tfbmin}
\end{equation}
which is much smaller than the typical timescales considered in our
analysis, but is useful to bear in mind when considering the discussion
in Sec.~\ref{sec:slope}.

\begin{figure*}
  \center
  \includegraphics[height=0.30\textwidth, angle=-90]{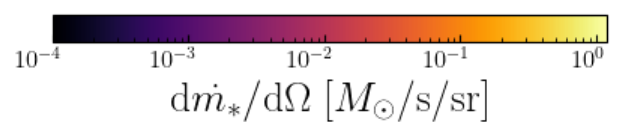}
  \hspace{1.0cm}
  \includegraphics[height=0.30\textwidth, angle=-90]{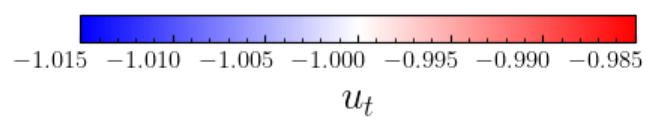}\\
  {\large \texttt{SFHO-q1.0}}\vspace{0.20cm}  \\
  \includegraphics[width=0.32\textwidth]{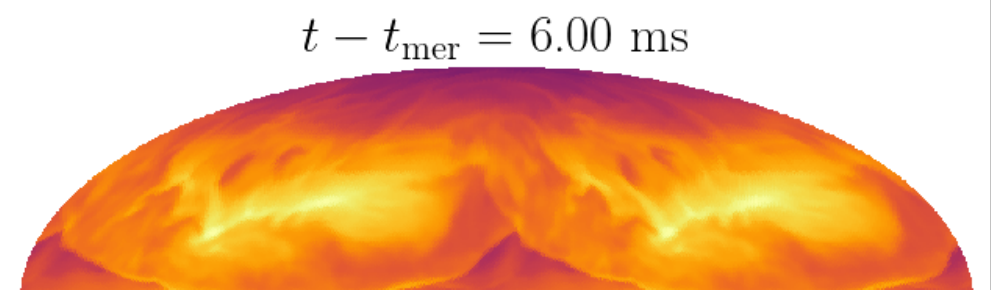}
  \includegraphics[width=0.32\textwidth]{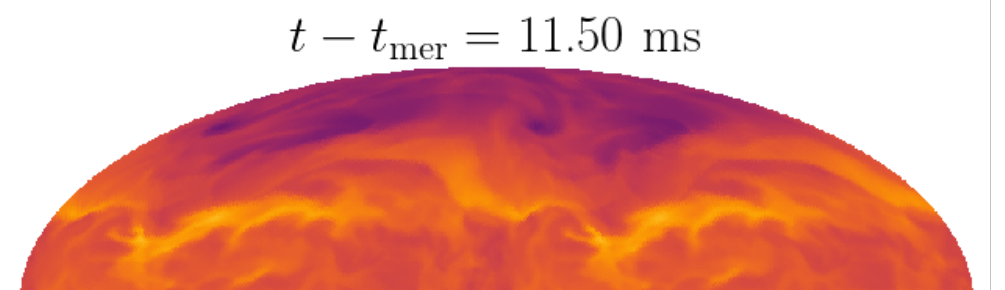}
  \includegraphics[width=0.32\textwidth]{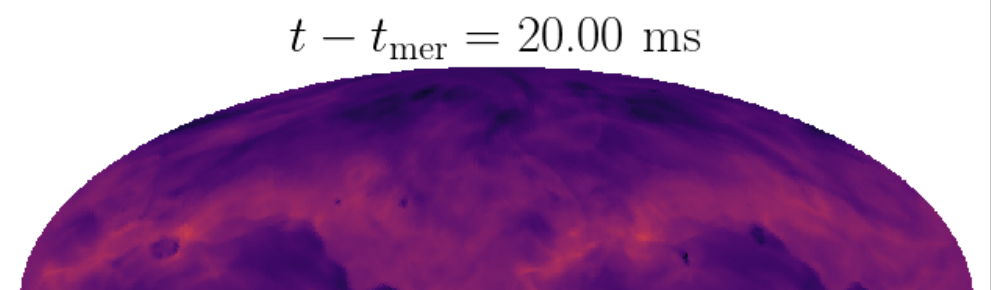}\\
  \includegraphics[width=0.32\textwidth]{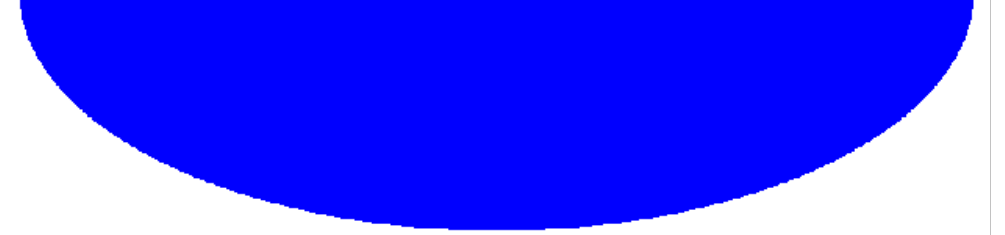}
  \includegraphics[width=0.32\textwidth]{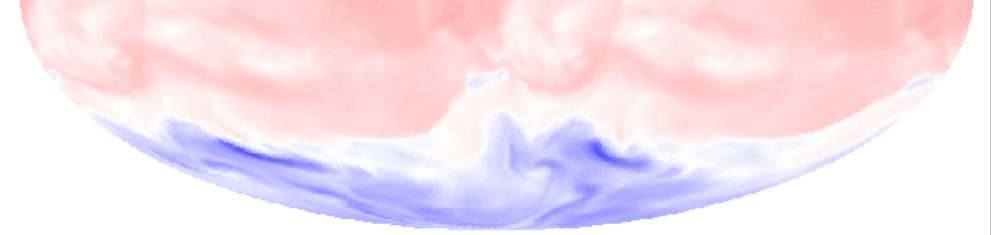}
  \includegraphics[width=0.32\textwidth]{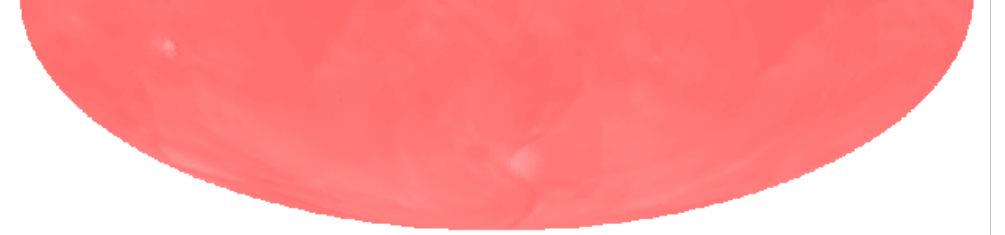}\\
  \vspace{0.5cm}{\large \texttt{SFHO-q.75}}\vspace{0.20cm}  \\
  \includegraphics[width=0.32\textwidth]{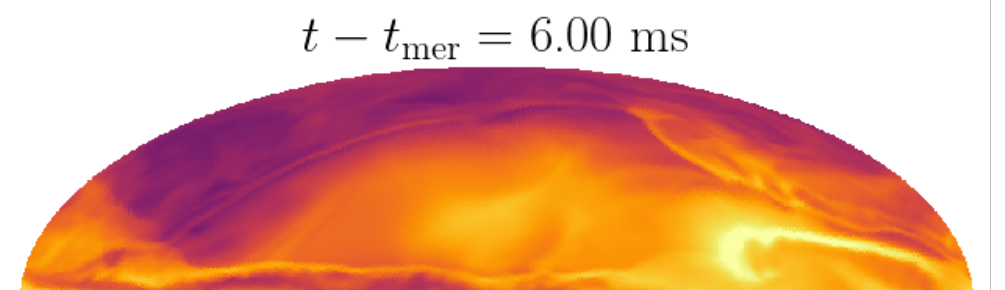}
  \includegraphics[width=0.32\textwidth]{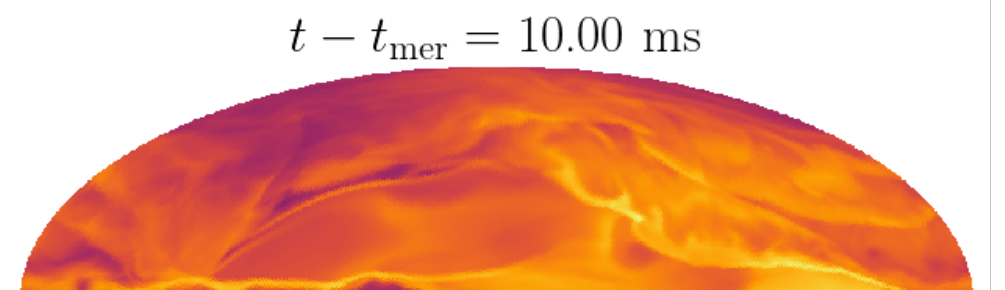}
  \includegraphics[width=0.32\textwidth]{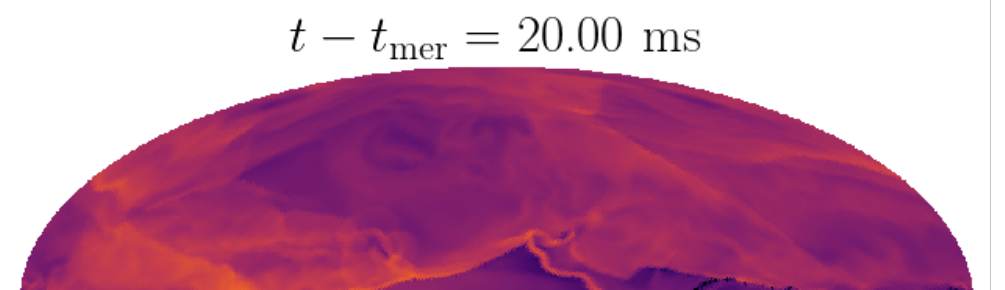}\\
  \includegraphics[width=0.32\textwidth]{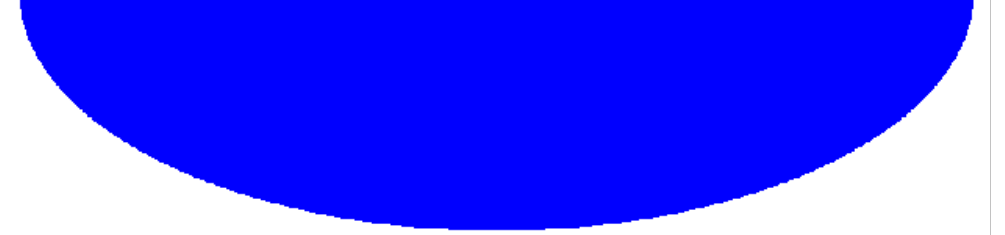}
  \includegraphics[width=0.32\textwidth]{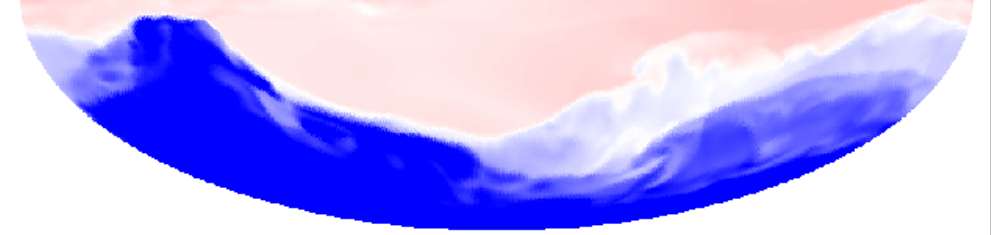}
  \includegraphics[width=0.32\textwidth]{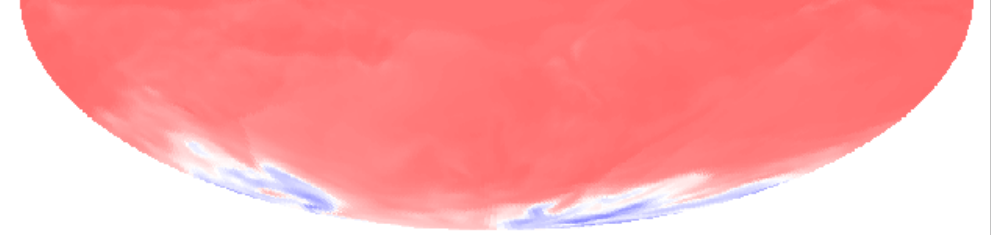}
  \caption{Snapshots of the material flowing through the detector surface
    at $R_d = 300\,M_\odot$ at various times of the simulation. Upper
    panels: \texttt{SFHO-q1.0} system. Lower panels: \texttt{SFHO-q.75}
    system. Upper hemispheres: mass flux per unit solid angle. Lower
    hemispheres: $u_t$ of the flowing material. The three columns
    correspond to different times during the evolution, as reported in
    Fig.~\ref{fig:fig3}.}
\label{fig:fig1}
\end{figure*}

What still needs to be determined is the mass $M_{\rm rem}$ of the
remnant, which we obtain with two different methods providing comparable
results. More specifically, in the case where a black hole forms within
the simulation timescale, we determine the black-hole mass using the
quasi-local integral on the apparent horizon surface as determined by the
vanishing of a null geodesic congruence expansion~\citep{Thornburg95,
  Dreyer02a}. In the cases where the remnant is a metastable HMNS, we
rely on the gravitational field measured at the detector surface. In
particular, in a 3+1 decomposition of spacetime~\cite[see,
  \eg][]{Gourgoulhon2007, Rezzolla_book:2013}, the covariant $tt$ metric
component is given by
\begin{equation}
g_{tt} = - \alpha^2 + \beta_i \beta^i 
\end{equation}
where $\alpha$ and $\beta^i$ are the lapse function and shift vector,
respectively. On the other hand, in the weak-field approximation of
general relativity, such a metric function at the detector
surface\footnote{We have verified that time-variation of $g_{tt}$ at the
detector surface is negligible so that it is reasonable to consider the
spacetime as stationary there.} can be related to the Newtonian
gravitational potential $\Phi$ as
\begin{equation}
g_{tt} \sim -1 - 2 \Phi = \frac{2M_{\rm rem}}{R_d} - 1 \,,
\label{eq:pot}
\end{equation}
so that
\begin{equation}
  M_{\rm rem} \simeq \frac{R_d}{2}\left(g_{tt} +1\right) \,.
\label{eq:pot_2}
\end{equation}

Equation~\eqref{eq:pot_2} thus provides an independent measurement of
the mass relevant for the fallback dynamics, including, in the case of
a black-hole formation, the eventual accretion disk, though it is
negligible in the cases studied here. The gravitational mass measured
via Eq.~\eqref{eq:pot_2} this method agree almost exactly with the ADM
mass in the simulation domain. By applying these two methods, we find
their results to agree to within $2\%$ for the two binaries where a
black hole forms (see Tab.~\ref{tab:1}). For convenience, since it can
be used equally well in the case of a black hole or HMNS, we will use
we the remnant mass computed via Eq.~\eqref{eq:pot_2} for all of the
binary systems.

As a concluding remark, we note that to avoid over-sampling the polar
region of our spherical detectors while only having a few cells near the
equator, as would be the case when using spherical coordinates to
represent the discrete grid on the 2-sphere, we developed a code where
the surface of the detector is are discretized following the
\texttt{HEALpix}
scheme~\citep{Gorski2005}\footnote{\href{https://healpix.sourceforge.io}{https://healpix.sourceforge.io}}. In
this coordinate system, all the cells have the same surface area and a
lower overall resolution is sufficient to capture the relevant
characteristics of the ejected matter as it crosses the detector.

%===================================================================
\subsection{Dynamics of the bound ejecta}
\label{sec:slope}
%===================================================================

\begin{figure*}
  \center
  \includegraphics[height=0.30\textwidth, angle=-90]{fig1_cb1.pdf}
  \hspace{1.0cm}
  \includegraphics[height=0.30\textwidth, angle=-90]{fig1_cb2.pdf}\\
  {\large \texttt{TNTYST-q1.0}}\vspace{0.20cm}  \\
  \includegraphics[width=0.32\textwidth]{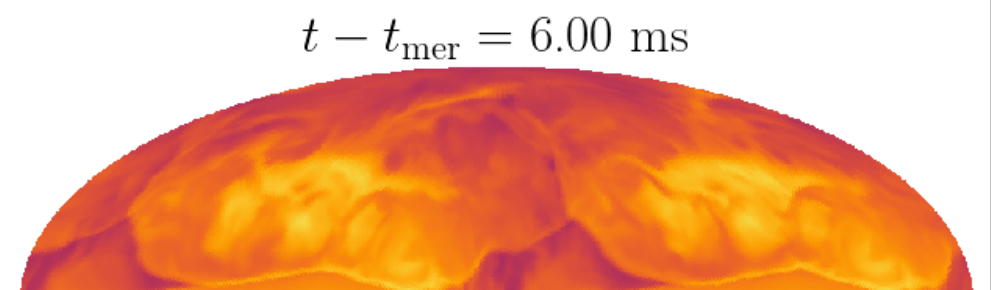}
  \includegraphics[width=0.32\textwidth]{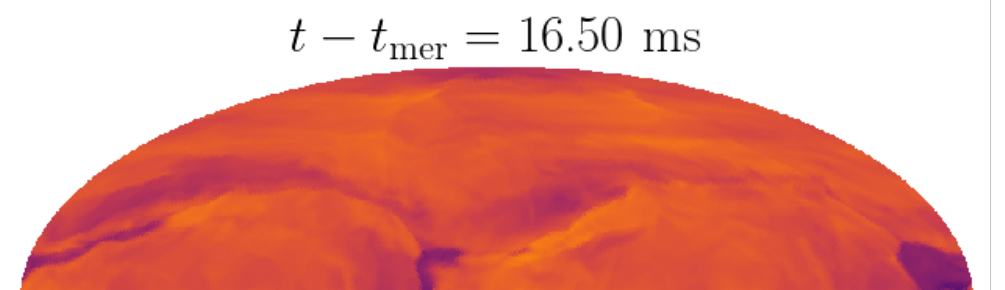}
  \includegraphics[width=0.32\textwidth]{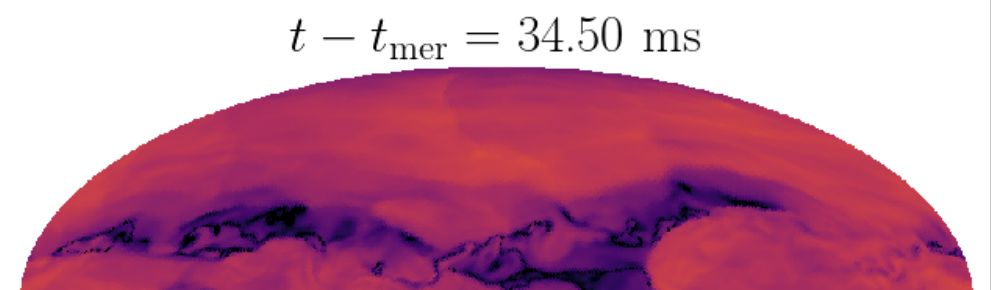}\\
  \includegraphics[width=0.32\textwidth]{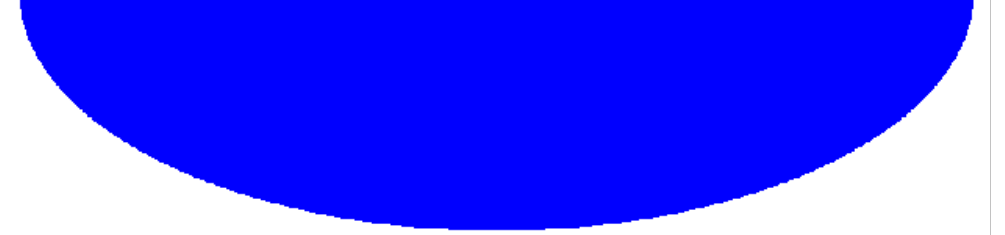}
  \includegraphics[width=0.32\textwidth]{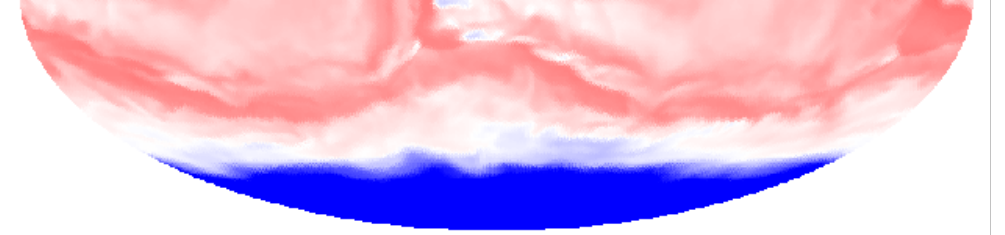}
  \includegraphics[width=0.32\textwidth]{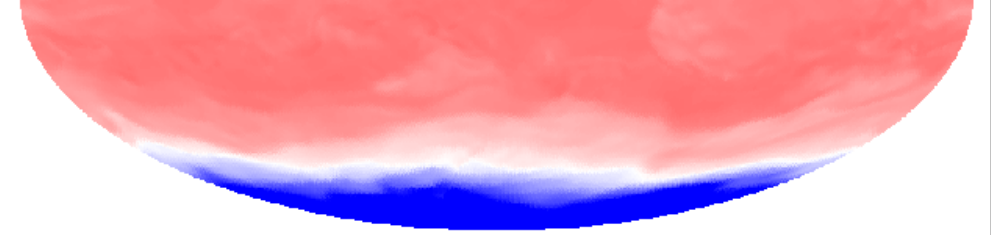}\\
  \vspace{0.5cm}
  {\large \texttt{TNTYST-q.75}}\vspace{0.20cm}\\
  \includegraphics[width=0.32\textwidth]{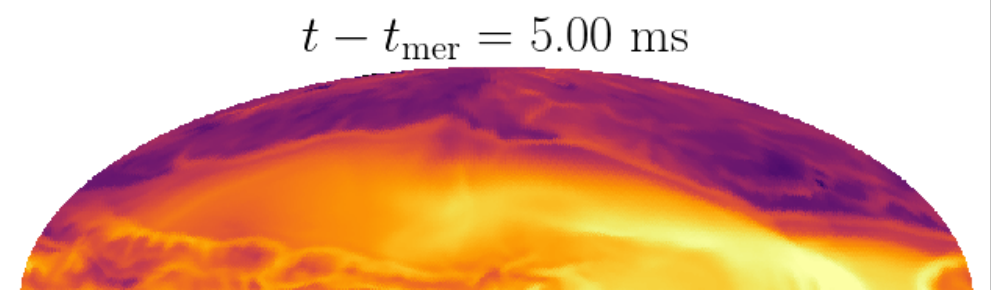}
  \includegraphics[width=0.32\textwidth]{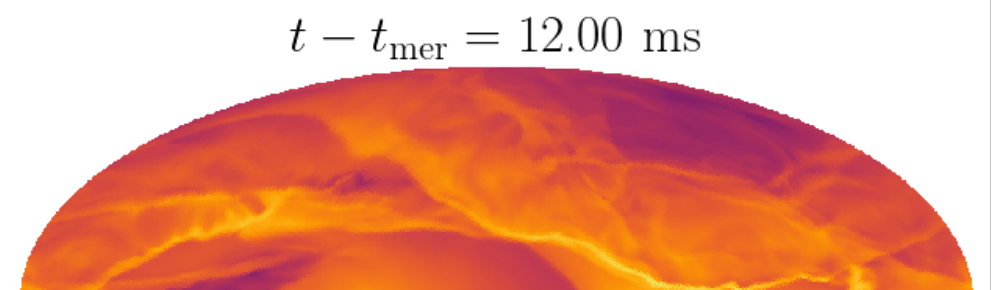}
  \includegraphics[width=0.32\textwidth]{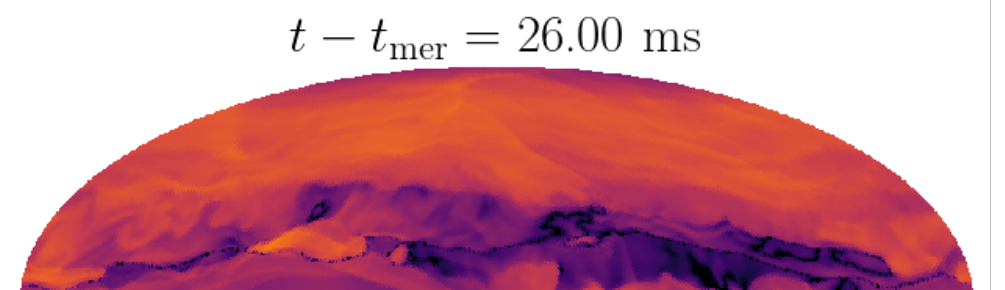}\\
  \includegraphics[width=0.32\textwidth]{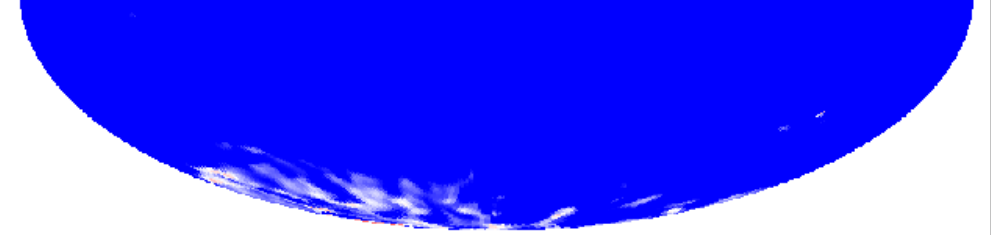}
  \includegraphics[width=0.32\textwidth]{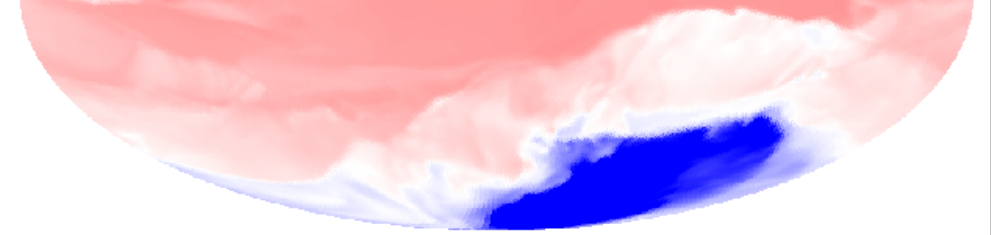}
  \includegraphics[width=0.32\textwidth]{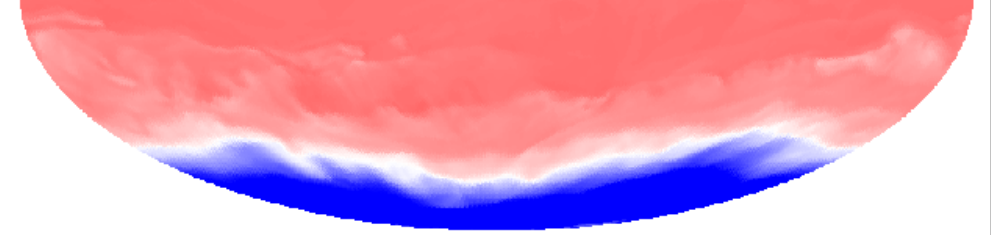}
  \caption{Same as Fig.~\ref{fig:fig1}, but for the two \texttt{TNTYST}
    binaries.}
  \label{fig:fig2}
\end{figure*}

In Fig.~\ref{fig:fig1}, we present snapshots at representative times of
the properties of the material crossing the detector sphere for the two
SFHO binaries. For each column, the top hemispheres report the
instantaneous rest-mass flux per unit solid angle $d\dot{M}_*/d\Omega$,
while the bottom hemispheres show the values of $u_t$, with red (blue)
shadings marking matter that is gravitationally bound (unbound). The
colour bar for $u_t$ was chosen to easily distinguish the bound material
in red from the unbound material in blue. Above each panel, we report the
times of the snapshots relative to the merger time, which, as customary,
is defined as the coordinate time at which the amplitude of the
$\ell=m=2$ component of the gravitational-wave strain reaches its first
maximum. In Fig.~\ref{fig:fig2}, we present the same data for the two
TNTYST binaries.

While the properties of the ejected material in BNS mergers has been
discussed in a number of papers~\citep[see, \eg][]{Sekiguchi2015,
  Bovard2017, Most2020e, Camilletti2022, Papenfort:2022ywx},
Figs.~\ref{fig:fig1} and~\ref{fig:fig2} show rather clearly a feature
that has not been remarked sufficiently so far. More specifically, and
irrespective of the mass ratio of the binary, the ejected material is
initially mostly unbound (left column) and then the fraction of bound
ejecta progressively increases (middle column) until a stage is
reached where the ejecta is predominantly (right column). In addition,
the geometrical distribution of the ejected matter is such that the
front of bound material progresses from the equatorial regions up to
the polar directions, ending up by filling the whole sphere. Finally,
while the early unbound ejecta is strongly anisotropic as a result of
the tidal tails, the bound ejecta is quite isotropic. By comparing the
upper and lower panels for each column, we find that the mass
asymmetry in the binary increases the anisotropy in the polar
direction of the early unbound ejecta (right column), as expected from
its origin from the tidal tails, retaining such a property also at
later time (right column). A more careful analysis of the polar
distribution of the bound ejecta (not shown in Figs.~\ref{fig:fig1}
and~\ref{fig:fig2}) reveals that the anisotropies are smaller than
$15\%$ for $t-t_{\rm mer} \gtrsim 20\,{\rm ms}$, i.e., for the bound
matter the mass flux differs by less than $15 \%$ across radial
directions at these times. Therefore, the bound ejecta can be
considered as essentially isotropic at late times.

By contrasting Figs.~\ref{fig:fig1} and~\ref{fig:fig2} we can appreciate
the impact of the EOS and in particular the role played by the stiffness
(the SFHO EOS is stiffer than the TNTYST). More specifically, when
considering a softer EOS in Fig.~\ref{fig:fig2}, we find that the
transition from unbound to bound is still present, however to a lesser
extent, where significant unbound flux is found even a later times in the
high-latitude regions. Similarly, the distribution of bound material at
intermediate and late times (middle and right columns) is less isotropic,
with a clear presence of unbound material at very high latitudes as a
result of a collimated outflow in the case of the TNTYST EOS. The origin
of the slightly different behaviour is to be found in the different
nature of the remnant, which is a black hole in the case of the
\texttt{SFHO} EOS, while it is a HMNS in the case of the TNTYST EOS, that
is metastable at least over the time window of the simulations (see
Tab.~\ref{tab:1}). Under these conditions, and as shown in a number of
recent works~\citep[see \eg][]{Fujibayashi2017, Fujibayashi2023}, the
HMNS is responsible for a neutrino-driven, collimated and mildly
relativistic matter outflow, which is clearly visible in the blue polar
regions in the middle and right columns of Fig.~\ref{fig:fig2}. Clearly,
a similar behaviour can be found when analysing the neutrino fluxes in
the two sets of binaries. Overall, what Figs.~\ref{fig:fig1}
and~\ref{fig:fig2} help appreciate rather transparently is that the
ejection of unbound material is active as long as a stable merger remnant
is present and is confined in the high-latitudes region. However, such
unbound outflow is rapidly shutdown as soon as a black hole is formed and
when this happens the ejected matter is mostly bound and with an
essentially isotropic distribution. These considerations apply in the
same manner to equal and unequal-mass binaries and quite independently of
the EOS at least when considering the different nature of the remnant.

\begin{figure*}
  \center
  \includegraphics[width=0.4\textwidth]{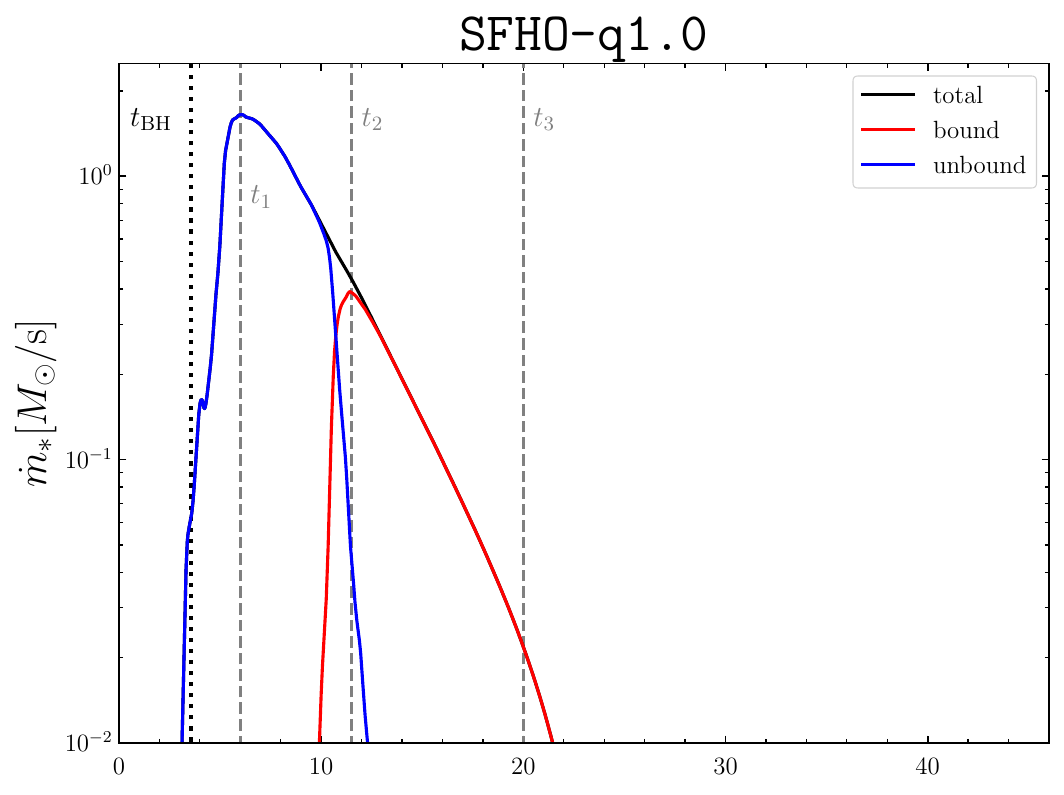}
  \hspace{0.5cm}
  \includegraphics[width=0.38\textwidth]{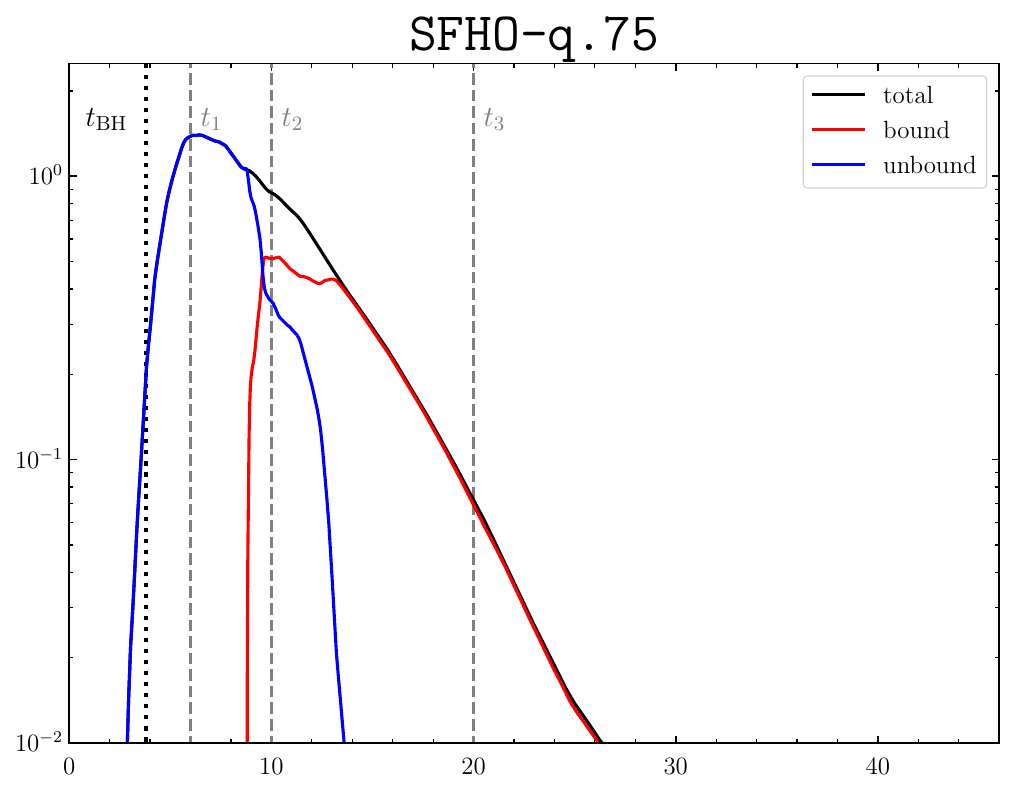}
  \includegraphics[width=0.4\textwidth]{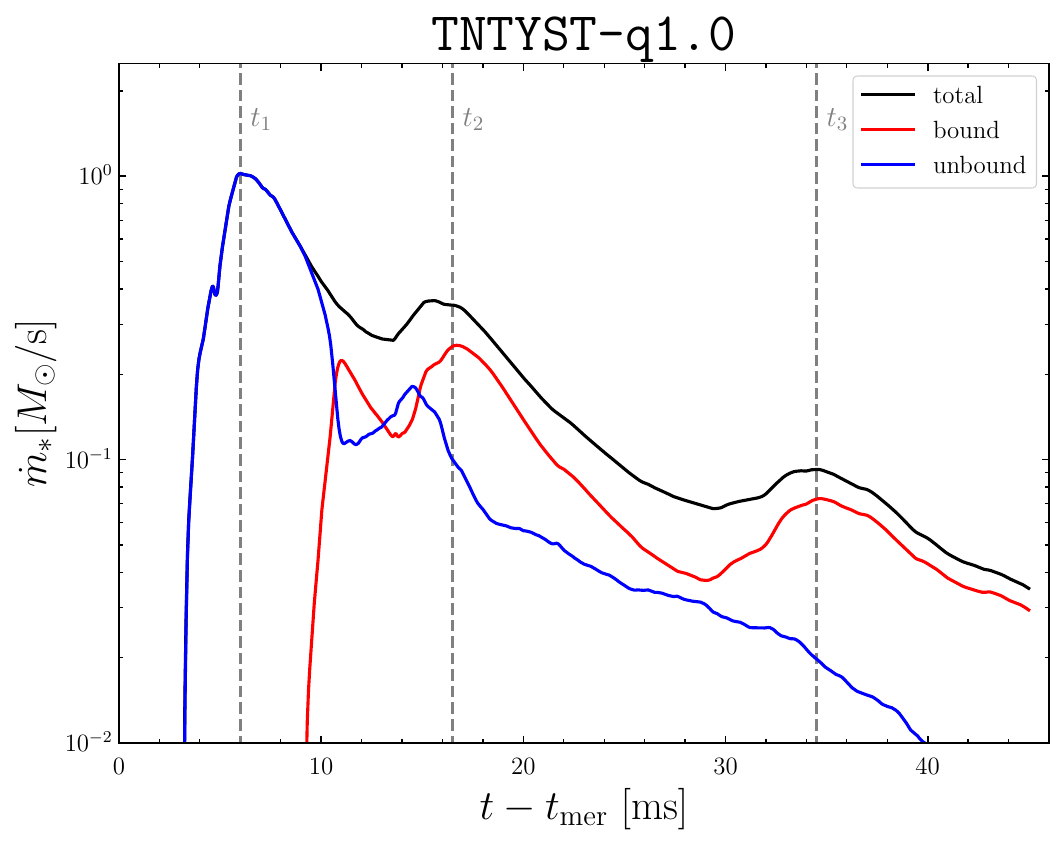}
  \hspace{0.5cm}
  \includegraphics[width=0.38\textwidth]{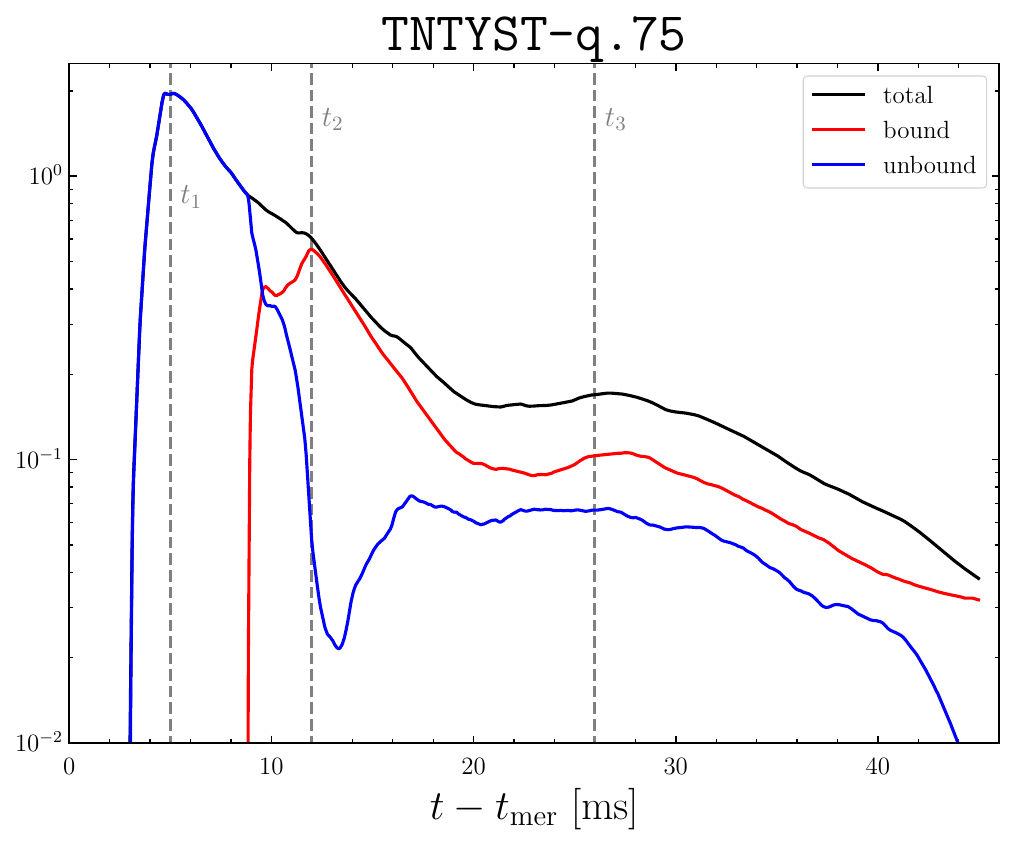}
\caption{Mass-ejection history of the four simulations. All times are
  measured from the merger time, and the mass flux is separated in bound
  (red), unbound (blue) and total mass flux (black). For all binaries,
  the vertical grey dashed lines mark the moments where the snapshots of
  Figs.~\ref{fig:fig1} and \ref{fig:fig2} are taken. For the two SFHO
  binaries (upper panels), the time of black-hole formation is also
  marked, with a black dotted line.}
\label{fig:fig3}
\end{figure*}

Figure~\ref{fig:fig3} provides a more quantitative history of the mass
ejection at the detector of our four binaries as measured, distinguishing
the total (black solid line) rest-mass flux from the flux of bound (red
solid line) and unbound material (blue solid line). The grey solid lines
report the times of the three snapshots presented in Figs.~\ref{fig:fig1}
and~\ref{fig:fig2}, thus allowing a clear identification over the
ejection history and highlighting that they correspond respectively to the
maxima of the total and bound rest-mass fluxes, and of the late-time
behaviour. Furthermore, in the case of the SFHO binaries (top
row), vertical black solid lines mark the time of black-hole formation.

Figure~\ref{fig:fig3} also allows one to appreciate in great detail the
general behaviours discussed above and relative to Figs~\ref{fig:fig1}
and~\ref{fig:fig2}. More precisely, it is possible to note that for the
two SFHO binaries (top row) a sharp transition from unbound to bound
ejection appears at about $10\,{\rm ms}$ post-merger, thus pointing out
that the episode of unbound ejection is rather short-lived and restricted
to the phase of dynamical ejection. Soon after the remnant collapses to a
black hole at $t-t_{\rm mer} \sim 4\,{\rm ms}$, in fact, the rest-mass
flux drops significantly, leaving only a small disk with mass ratio
$M_{\rm disk}/M_{\rm rem} \simeq 0.02$ that will be responsible for the
subsequent secular mass ejection. Indeed, this is a very small disk and
this is most likely due to the fact that in these binaries the black hole
is produced very rapidly after merger and the angular momentum
redistribution in the merger remnant has not taken place yet, hence with
most of the rest-mass on unstable circular orbits that will be rapidly
accreted once the black hole is produced. We expect that binaries with
the same SFHO EOS but slightly smaller chirp mass would lead to a longer
equilibrium of the HMNS and hence to larger discs once the black hole is
formed.

Note that in our two SFHO binaries, the disk produced is very small due
to the prompt collapse of the remnant. We conjecture that this small mass
inhibits the ejection of unbound material from the disk, which would be
due to MHD processes or neutrino-driven processes (which in any case
would be very sub-dominant, see comparison in \citealt{Gill2019}) in a
larger disk that could sustain a higher pressure, as seen, \eg
by~\citet{Fahlman2018, Fujibayashi2018}. The low mass of the disk in our
SFHO cases only allows for bound ejecta, as shown
in~Fig.~\ref{fig:fig3}. Note also that the transition from unbound to
bound ejecta suggests that the fallback material will always remain
\textit{below} the unbound matter, with a somewhat precise geometrical
separation. Indeed, inspecting the velocities of the bound and unbound
components and extrapolating in time leads us to conclude that the two
components will remain distinct also in the subsequent evolution. This
behaviour has important consequences that we will discuss in more detail
in Sects.~\ref{sec:2} and \ref{sec:discussion}.

\new{Interestingly, this interplay between the mass of the disk and the
  ratio of the bound to unbound ejected material may suggest an
  anti-correlation between the energy in the prompt GRB emission and that
  in the extended emission. Indeed, one would expect more massive disks
  to be able to sustain a powerful jet for a longer time [before the disk
    either depletes or enters a magnetically arrested state, leading in
    both cases to a decrease in the jet power~\citep{Gottlieb:2023b}] and
  thus to a larger energy release in the prompt emission. On the other
  hand, a less massive disk would lead to comparatively larger amount of
  bound material being ejected and therefore to a larger energy reservoir
  for the extended emission when compared to the
  prompt GRB signal.}

When considering instead the evolution of the ejected material for the
TNTYST binaries (bottom row in Fig.~\ref{fig:fig3}), it is possible to
recognise the same features discussed for the \texttt{SFHO} binaries with
the important difference that the ejection of unbound matter does not
drop rapidly since no black hole is formed in this case during the
simulation. As a result, the HMNS is able to eject, especially in the
polar region, a neutrino-driven, collimated and mildly relativistic
outflow of unbound matter. Such an unbound flow, however, is essentially
all the time smaller than that of the bound material, which represents
the dominant contribution to the matter crossing the detector. Note also
that the bound flow does not exhibit a monotonically decreasing evolution
with time and has episodes in which its rate increases although of a
factor of a few only (see, \eg $t-t_{\rm mer} \sim 35\,{\rm ms}$ for the
\texttt{TNTYST-q1.0} binary and $t-t_{\rm mer} \sim 35\,{\rm ms}$ for the
\texttt{TNTYST-q.75} binary). Given the very complex dynamics of the
HMNS, it is difficult to disentangle the origin of these
``rebrightening'' episodes, which we attribute to the material being
ejected after being shock-heated by the bound material that in the
meanwhile has fallen back onto the HMNS. As we will show later, the
fallback time of this late-time bound ejecta is very short and hence will
have no impact on the EM signatures of the binary. The situation would be
rather different if this rebrightening material had longer fallback
times, as we discuss in Appendix.~\ref{sec:B}.

Finally, Tab.~\ref{tab:1} reports not only the total masses ejected, but
also the corresponding components that are in bound and unbound
matter. Note that quite independently of the EOS, unequal-mass binaries
tend to have larger values of ejected mass; on the other hand, the softer
TNTYST EOS favours bound ejection by almost a factor of two in the
equal-mass case. \new{We have also estimated that, quite generically,
  about $10^{-6}\,M_{\odot}$ of bound ejecta have a fallback time larger
  than 100 seconds; this mass increases by a factor of about five if we
  consider a fallback time larger than 10 seconds}.

\begin{figure}
\includegraphics[width=0.95\columnwidth]{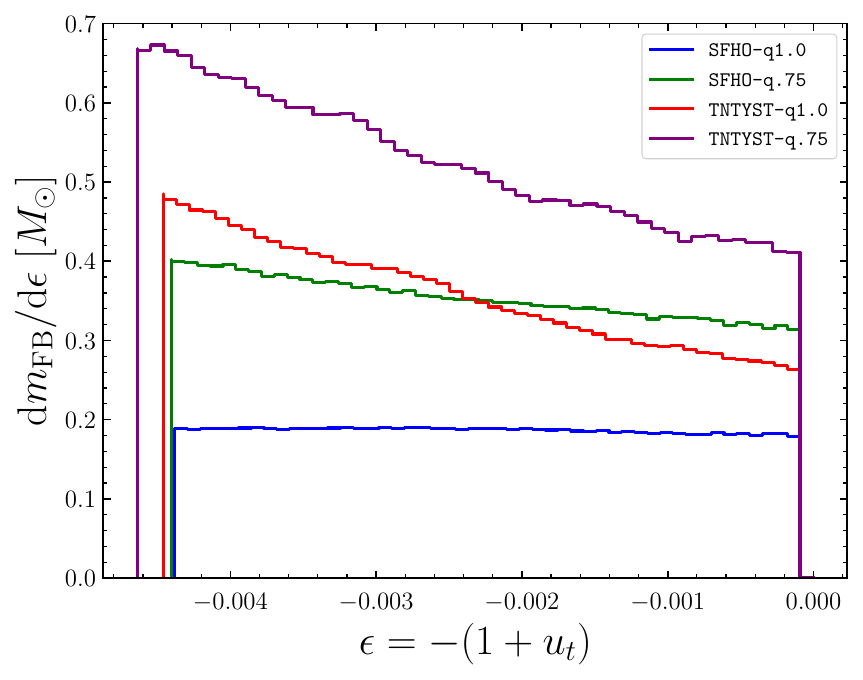}
\caption{Mass histogram of the orbital energy in the bound ejecta counted
  over the entire simulation. For explanations on the lower bounds of
  these histograms, report to Sec.~\ref{sec:method}.}
\label{fig:fig4}
\end{figure}

%===================================================================
\subsection{Dynamical properties of the fallback accretion}
\label{sec:eos}
%===================================================================

We next concentrate on illustrating the general characteristics of the
significant amount of fallback inflow and discuss how it can lead to a
long-term EM emission. We start by presenting in Fig.~\ref{fig:fig4} the
distribution of the orbital specific energy $\epsilon:=-(1+u_1)$ in the
fallback matter for our four binaries. Since $\epsilon = 0$ corresponds
to marginally bound matter, we obviously consider the distributions of
bound matter only for $\epsilon < 0$, the with lower bounds ($\epsilon
\simeq -0.0045$) being set by the tightest orbits we can measure at the
detector [see the discussion around Eq.~\eqref{eq:tfbmin}].

Remarkably, these distributions are essentially flat near the $\epsilon
\lesssim 0$ bound and they remain nearly constant over the entire range
of specific energies for the SFHO binaries (blue and green solid lines).
This same can be said for the binaries with the TNTYST EOS (red and
purple solid lines in Fig.~\ref{fig:fig4}). The fact that these
distributions are essentially flat has an important implication since, as
already discussed early on in the study of fallback of debris from TDEs,
\citep[see, \eg][]{Rees1988, Evans1989, Phinney1989}, in this case it is
not difficult to show that the fallback accretion rate $\dot{m}_{\rm FB}$
must follow a power-law in time, namely, using Eq.~\eqref{eq:orb} it is
trivial to obtain that
\begin{equation}
\label{eq:powerlaw}
  \dot{m}_{\rm FB} := \frac{d m_{\rm FB}}{d t_{\rm FB}} = \frac{d
    m_{\rm FB}}{d \epsilon} \frac{d \epsilon}{d t_{\rm FB}} \propto
  \frac{d m_{\rm FB}}{d \epsilon} \times t_{\rm FB}^{-5/3} \propto
  t_{\rm FB}^{-5/3} \,,
\end{equation}
The importance of the result~\eqref{eq:powerlaw} stems not only from the
very simple hypotheses employed to obtain it, but also because it has
been indirectly confirmed by some late-time observations of
TDEs~\citep[see, \eg][for a review]{Komossa2015}.

\begin{figure}
\includegraphics[width=0.95\columnwidth]{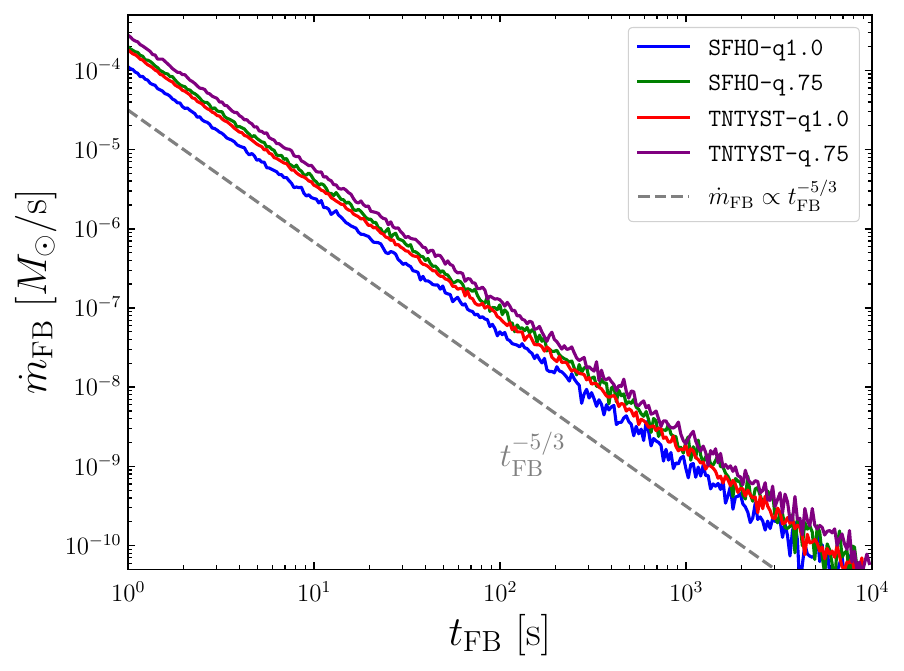}
\caption{Fallback accretion rate for the bound material. We present the
  four simulated binaries in solid lines, and an accretion rate with
  $\dot{m_{\rm FB}} \propto t_{\rm FB}^{-5/3}$ to guide the eye (dashed
  line). }
\label{fig:fig6}
\end{figure}

Interestingly, the power-law behaviour predicted by
Eq.~\eqref{eq:powerlaw} is exhibited also by our fallback material, which
obviously includes, in addition to the tidal ejecta, also the bound
material coming from the neutrino-driven and magnetically driven winds
from the merger remnants. More specifically, using the distribution of
fallback times, we can compute the fallback accretion rate as a function
of time and report the results in Fig.~\ref{fig:fig6}, where we also plot
a $t^{-5/3}$ power-law decay to guide the eye. As anticipated, for all
the cases considered, and hence rather independently of the EOS or the
mass ratio, the fallback accretion proceeds with a power-law decay
$\dot{m}_{\rm FB} \propto t^{-5/3}$ (see black dashed line). This result,
which was already known for neutron-star-black-hole binaries, either in
Newtonian calculations \citep[\eg SPH][]{Rosswog2007_fb} or in full
general relativity \citep{Chawla:2010sw, Kyutoku2015}, is now confirmed
also from rather generic conditions of BNS mergers. Figure~\ref{fig:fig6}
also highlights that the power-law result for the fallback accretion rate
depends only very weakly on the specific-energy distributions. As shown
in Fig.~\ref{fig:fig4}, in fact, the specific-energy distributions of our
TNTYST binaries are only approximately constant, showing instead an
excess (deficit) at lower (higher) specific energies. Yet, this does not
affect the $\sim t^{-5/3}$ behavior of the fallback accretion rate of the
TNTYST binaries. We believe this is because material with low specific
energy will affect only the onset of the fallback process; however, the
long-term features of the fallback accretion are essentially determined
by the matter that is only marginally bound and hence with $\epsilon
\lesssim 0$. Because all of the distributions in Fig.~\ref{fig:fig6} are
essentially constant at $\epsilon \lesssim 0$, it is not surprising that
all of our binaries reproduce the $t^{-5/3}$ behaviour irrespective of
their EOS, mass ratio, or nature of the merger remnant. Note also from
Fig.~\ref{fig:fig6} that the measured accretion rates between $t_{\rm FB}
\sim 10^2-10^3\,{\rm s}$ are of the order $\dot{m}_{\rm FB} \sim
10^{-9}-10^{-7}\,M_{\odot}/{\rm s}$. As a result, using a crude
conversion of accretion rate to luminosity, \ie $L_{\rm acc} \sim 0.1
\times \dot{m}_{\rm FB}c^2$, it is straightforward to obtain a very rough
estimate of the luminosities produced by this inflow, namely, $L \simeq
10^{44}-10^{44}\,{\rm erg/s}$. In Sec.~\ref{sec:3} we will confirm and
refine this result using a more detailed EM model.

\begin{figure*}
  \center
  \includegraphics[width=0.45\textwidth]{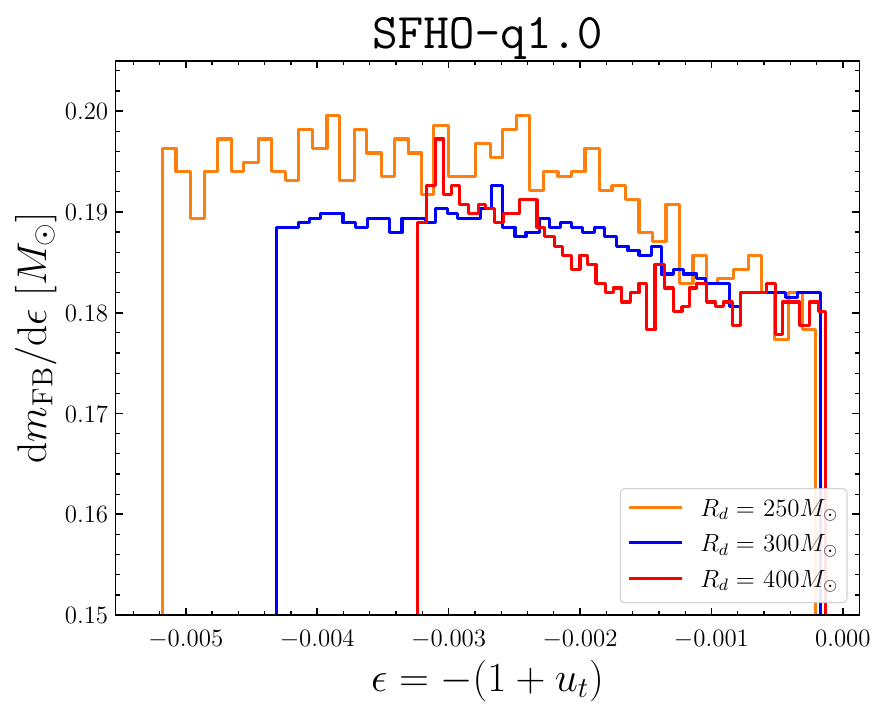}
  \hspace{0.5cm}
  \includegraphics[width=0.45\textwidth]{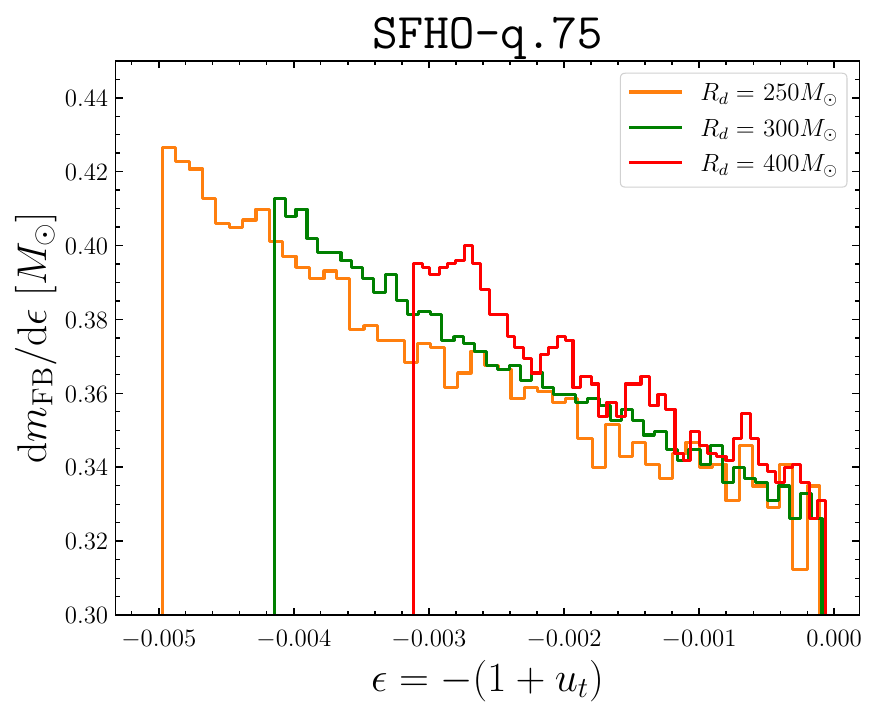}
  \caption{Comparison of mass distributions of orbital energy as measured
    for different detector positions in the domain. The histogram is
    calculated as in Fig.~\ref{fig:fig4}, by placing the detector at
    different radii, as per the legend. Left: \texttt{SFHO-q1.0}
    system. Right: \texttt{SFHO-q.75} system. }
\label{fig:fig5}
\end{figure*}

The robustness of the result presented in Fig.~\ref{fig:fig6} is also
underlined by the two panels in Fig.~\ref{fig:fig5}, where we show the
same distributions as in Fig.~\ref{fig:fig4} for the SFHO binaries, but
when computed at detectors that are either at larger (\ie
$400\,M_{\odot}$) or smaller (\ie $250\,M_{\odot}$) radii. In these
cases, the lower bounds in the specific-energy distributions obviously
change, increasing (decreasing) as the detector radius is decreased
(increased) simply because of the differences in the tightest orbits that
the detectors can capture. Clearly, the overall distributions do not show
significant changes with the location of the detector and, more
importantly, they are essentially identical at the highest specific
energies, \ie at $\epsilon \simeq 0$. As a result, following the logic
discussed above, the corresponding fallback accretion rates all yield the
same $\sim t^{-5/3}$ power-law behaviour (not shown here). Furthermore,
since the specific-energy distributions all agree on around $\epsilon
\simeq 0$, we can conclude that the conversion of internal energy into
kinetic energy is not unbinding material significantly between the
different radii, hence justifying our use of $u_t$ over $hu_t$ to
determine the bound material.

\begin{figure*}
  \center
  \includegraphics[width=0.1\textwidth,angle=-90]{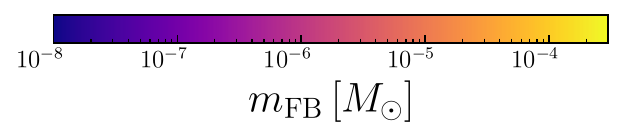}\\
  \includegraphics[width=0.450\textwidth]{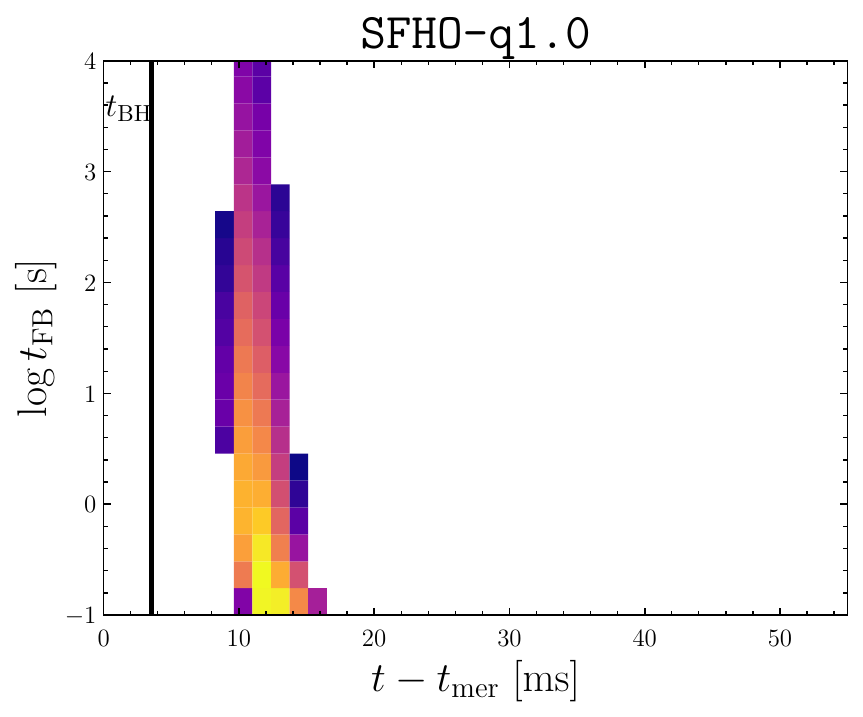}
  \hspace{0.5cm}
  \includegraphics[width=0.425\textwidth]{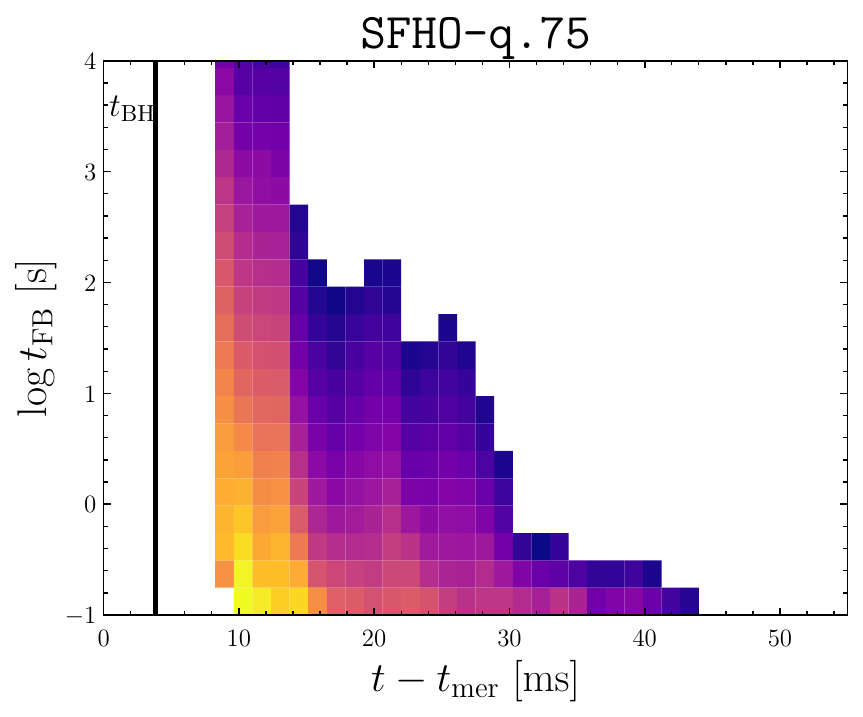}
  \includegraphics[width=0.450\textwidth]{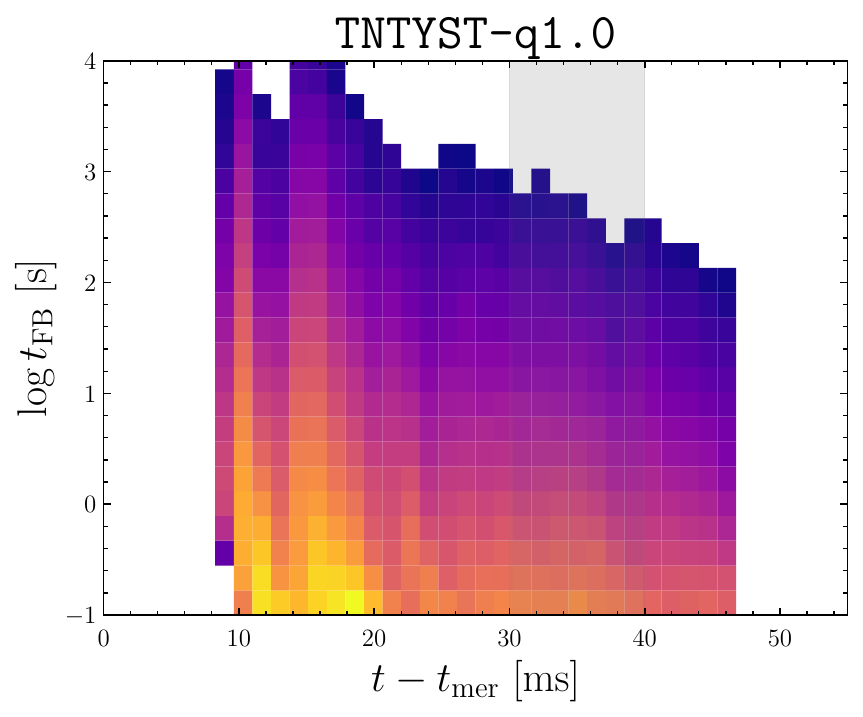}
  \hspace{0.5cm}
  \includegraphics[width=0.425\textwidth]{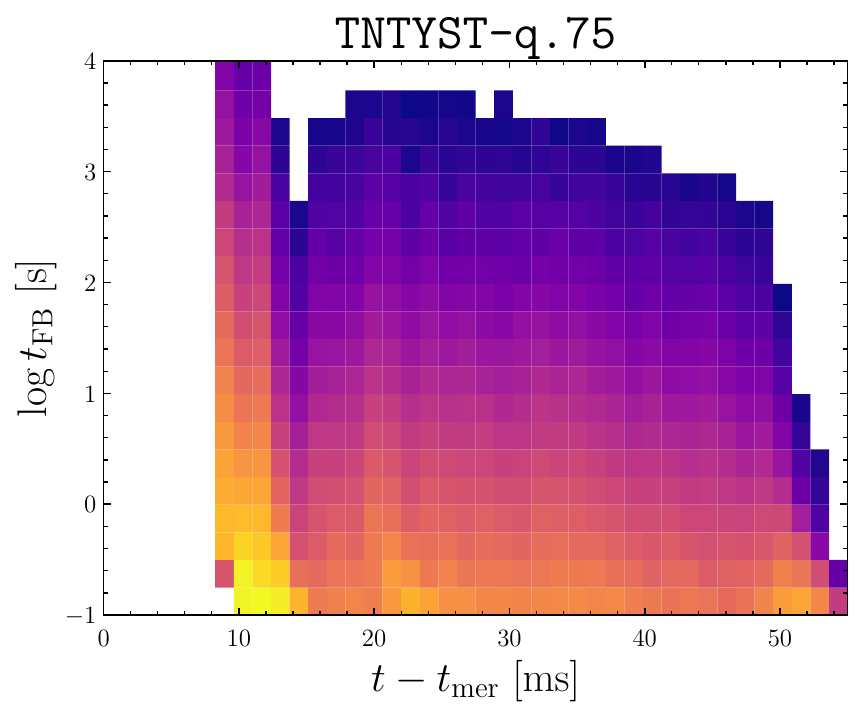}
  \caption{Two-dimensional mass histogram of bound matter as per its
    ejection time (x-axis) and fallback time (y-axis), for all four
    binaries studied. These histograms evidence the fallback time of the
    bound material ejected at different stages of the merger and
    post-merger phases. For the two SFHO binaries (top panels), we also
    indicate the time of formation of the black hole (vertical black
    line). For the \texttt{TNTYST-q1.0} system (lower left panel), we
    show the 10-ms segment that is reproduced in sequence to study the
    consequences of very long-term ejection (grey-shaded area; see
    discussion in Sec.~\ref{sec:B}).}
\label{fig:fig7}
\end{figure*}

Finally, in Fig.~\ref{fig:fig7} we present the distributions in time and
fallback time of the matter accreted onto the remnant for the four
binaries considered in our analysis. Note that in the first $10\,{\rm
  ms}$ after merger all binaries exhibit large ejections of bound matter
and with a broad distribution of fallback times, \ie from $10^{-1}\,{\rm
  s}$ up to $10^{4}\,{\rm s}$, although most of the bound mass has
$t_{\rm FB} \lesssim 10^{2}\,{\rm s}$. As the evolution proceeds,
however, differences brought in by the EOS and the mass ratio start to
emerge. In particular, in the case of the stiffer SFHO binaries and
equal-mass binaries, the bound ejecta are essentially suppressed after
$\lesssim 20\,{\rm ms}$ (see also top-left panel of Fig.~\ref{fig:fig3}),
while they continue to be present for the unequal-mass case but with a
much reduced variance in fallback times, so that $t_{\rm FB} \lesssim
10^{2}\,{\rm s}$ after about $20\,{\rm ms}$ after merger for
\texttt{SFHO-q.75}. On the other hand, in the case of the softer TNTYST
binaries, large amounts of bound ejecta are produced for much longer
timescales (\ie $t-t_{\rm mer} \lesssim 40\,{\rm ms}$) and with a broad
distribution of fallback times. The main difference in this case is to be
found in the mass ratio, with the ejection being essentially shut-down
for $t-t_{\rm mer} \lesssim 40\,{\rm ms}$ in the case of the equal-mass
binary, while being persistent for the unequal-mass \texttt{TNTYST-q.75}
binary. Interestingly, for the latter the distribution of fallback times
reduces significantly at late times (\ie $t_{\rm FB} \lesssim 10\,{\rm
  s}$ for $t-t_{\rm mer} \gtrsim 50\,{\rm ms}$), most likely because of
the cooling of the HMNS, which reduces the specific energy of the ejecta
in the neutrino and magnetically driven winds [see Eq.~\eqref{eq:orb}].
Nevertheless, one could expect that by running the simulation further,
provided that the HMNS remains stable against gravitational collapse, a
secondary, collimated and magnetically driven wind would emerge, has
shown by several recent works~\cite{Combi2023, Kiuchi2023}. The fact that
this wind does not appear in our simulation time might depend on several
factors: firstly, the physical mechanism at the basis of this ejection,
while argued to be likely tied to an $\alpha-\Omega$ dynamo, is still
poorly understood. Secondly, the resolution of our simulations is too low
to consistently resolve the magnetorotational instability in the outer
layer of the massive star, as well as in the high-density and
high-temperature regions of the disk, therefore delaying the onset of the
dynamo and resulting wind.

%-------------------------------------------------------------------
%-------------------------------------------------------------------
\section{Extended emission from fallback accretion}
\label{sec:3}
%-------------------------------------------------------------------
%-------------------------------------------------------------------

%===================================================================
\subsection{A semi-analytical model}
\label{sec:flow}
%===================================================================

In Sec.~\ref{sec:2}, we showed that significant fallback accretion of
material can occur on long timescales, \ie for $t_{\rm FB} \gtrsim
100\,{\rm s}$. The fallback rates reported in Fig.~\ref{fig:fig6} show
that this accretion is largely super-Eddington, such that the conversion
of the mass accretion rate $\dot{m}_{\rm FB}$ to the accretion luminosity
$L_{\rm acc}$ is not straightforward and requires a more precise
modelling than the crude one we have employed in the previous
section. Furthermore, the fallback material, just like the unbound
material, is subject to nucleosynthetic processes that contribute to the
generation of EM radiation that needs to be taken properly into
account. However, because the dynamics of the outflow-inflow cannot be
followed on such timescales by our simulations, a semi-analytical model
is needed in order to determine such a dynamics, from which a model for
the EM emission can be derived.

As indicated by the basic principles of orbital dynamics [see
  Eqs.~\eqref{eq:ecc} and~\eqref{eq:orb}] and by the inspection of the
numerical data from the simulations, material with long fallback times,
\eg $t_{\rm FB} \gtrsim 100\,{\rm s}$, has very eccentric
orbits. Furthermore, as noted in Sec.~\ref{sec:slope}, the ejection of
bound material is largely isotropic, such that the fallback dynamics can
also be considered independent of $\theta$ and $\phi$. Given these
considerations, it is reasonable to simplify our semi-analytic model by
by considering the inflow-outflow dynamics to be purely radial. In this
case, the properties of the fallback inflow reduce to the rest-mass
density and to the radial velocity fields $\rho$ and $v$ that are
functions only of time and radius $(r,t)$. However, since the electron
fraction of the material does depend on the latitude, we will account for
this in the determination of the fallback radiation in Sec.~\ref{sec:ee}.

To determine the dynamics of the infalling material we consider it to be
in radial free-fall. In addition, we consider all the fallback matter to
have been ejected at the same time, which marks $t = 0$ in our study of
the fallback dynamics. This assumption corresponds to neglecting the
duration of the ejection episode in the merger, which is $\Delta t_{\rm
  ej} \lesssim 100\,{\rm ms}$ (see Fig.~\ref{fig:fig7}), which is valid
for the fallback timescales considered. A fluid element from
freely-falling from an initial radius $R_i$ with zero velocity to a
radius $R_f$ will have a Newtonian free-fall time given by
\begin{equation}
\label{eq:tFF}
  t_{\rm FF}(R_i, R_f) = \sqrt{\frac{R_i^3}{2GM}} \left(
\sqrt{\frac{R_f}{R_i} \left( 1 + \frac{R_f}{R_i}\right)} +
\arccos\sqrt{\frac{R_f}{R_i}}\right) \,.
\end{equation}
and its  velocity at the final radius $R_f$ is simply
\begin{equation}
v_{\rm FF}(R_i, R_f) = \sqrt{2 GM \left( \frac{1}{R_f} -
  \frac{1}{R_i}\right)}\,,
\end{equation}

Consider now a radial segment of matter of size $d L_0$ initially at
radius $R_i$. After free-falling to radius $R_f<R_i$, this radial segment
will have spread out to a size $d L_f$ such that:
\begin{equation}
d L_f - d L_0 = v_{\rm FF}(R_i, R_f) \frac{\partial t_{\rm
    FF}}{\partial R_i}(R_i, R_f)d L_0 \,,
\label{eq:lf}
\end{equation}
since the trailing matter will take a time $t_{\rm FF}(R_i, R_f) - t_{\rm
  FF}(R_i - d L_0, R_f)$ to catch-up the leading matter, with velocity
$v_{\rm FF}(R_i, R_f)$ \new{(note that we here assume that the velocity
  vanishes at both edges simultaneously)}. By conservation of matter in
the free-fall, the densities $\rho_i$ and $\rho_f$ before and after the
free-fall are related by
\begin{equation}
\rho_i R_i^2 d L_0 = \rho_f R_f^2 d L_f \,,
\end{equation}
or, using Eq.~\eqref{eq:lf}
\begin{equation}
\label{eq:rho_prof}
  \rho_i = \rho_f \left(\frac{R_f}{R_i}\right)^2\left[ 1 + v_{\rm FF}(R_i,
  R_f)\frac{\partial t_{\rm FF}}{\partial R_i}(R_i, R_f)\right] \,.
\end{equation}

Hence, after setting $R_f=R_d$ and considering generic initial radius
$\bar{R}$ and time $\bar{t}$, Eq.~\eqref{eq:rho_prof} can be written as
\begin{eqnarray}
&&\rho(\bar{R}, \bar{t}) = \left(\frac{R_d}{\bar{R}}\right)^2 \rho(R_d, \bar{t} + t_{\rm
  FF}(\bar{R}, R_d)) \times \\
  && \hskip 2.0cm
  \times \left[ 1 + v_{\rm FF}(\bar{R}, R_d)\frac{\partial
    t_{\rm FF}}{\partial R_i}(\bar{R}, R_d)\right]\,,
\label{eq:sa}
\end{eqnarray}
which allows us to determine the density at an arbitrary radius $\bar{R}
> R_d$ and time $\bar{t} > 0$ in the flow from its value at the detector
$R_d$. As discussed in Sec.~\ref{sec:slope}, we know that at $R_d$ the
fallback accretion rate is
\begin{equation}
\dot{m}_{\rm FB}(t_{\rm FB}) = 4 \pi R_d^2 \rho(R_d, t_{\rm FB}) v_{\rm
  FF}(R_{\rm max}(t_{\rm FB}), R_d)\,,
\label{eq:rd}
\end{equation}
where $\dot{m}_{\rm FB}(t_{\rm FB}) \propto t_{\rm FB}^{-5/3}$ (\eg
Fig.~\ref{fig:fig6}) and where $R_{\rm max}(t_{\rm FB})$ is the maximum
radial coordinate reached by the ejected bound material before falling
back onto the remnant. In practice, $R_{\rm max}(t_{\rm FB})$ can be
computed from Eq.~\eqref{eq:tFF} after requiring that
\begin{equation}
t_{\rm FF}(R_{\rm max}(t_{\rm FB}), R_d) = \frac{t_{\rm FB}}{2}\,.
\label{eq:tff}
\end{equation}

Note that in our semi-analytic model, the density of fallback material is
zero for all $(r, t)$ such that:
\begin{equation}
r \geq R_d \left(1 + \frac{\sqrt{2 G M} t}{R_d} \right)^{2/3}\,,
\label{eq:marg}
\end{equation}
as the equality in~\eqref{eq:marg} corresponds to the orbit of material
with zero energy at infinity. Above the leading edge separating the
outgoing and ingoing fallback material (see dashed line in
Fig.~\ref{fig:cartoon}), the layer of unbound material will produce the
kilonova signal as it moves outwards (this is marked with a dashed line
in Fig.~\ref{fig:fig8}). As mentioned in Sec.~\ref{sec:slope}, the
velocities of the bound and unbound component of the ejecta are well
separated, such that we expect a gap between the flow of fallback
material and the unbound outflow. In Sec.~\ref{sec:discussion}, we
further discuss the consequences of this geometry on the observation of
radiation from the fallback component.

To solve for the density in the fallback inflow, we normalise the
fallback rate on the detector surface $\dot{m}_{\rm FB}$ using the total
mass of bound ejecta (Tab.~\ref{tab:1}), and then apply
Eqs.~\eqref{eq:lf}, \eqref{eq:sa}, \eqref{eq:rd}, and \eqref{eq:tff}. In
this way, it is possible to build a spacetime diagram of the
semi-analytic model for the infalling matter reporting the worldlines
(\ie the trajectories in spacetime) of shells of material with different
rest-mass densities (hence with different colours), and which we report
in Fig.~\ref{fig:fig8} for the representative \texttt{SFHO-q1.0} binary,
which has a total fallback mass of $1.60\times10^{-3}\,M_\odot$. Note
that the spacetime is reported in a double logarithmic scale and hence
timelike trajectories appear as spacelike, while the grey line near the
vertical axis represents the worldline of the detector's surface at $R_d
= 300\,M_\odot$, where the fallback accretion flow with $\dot{m}_{\rm FB}
\propto t^{-5/3}$ is enforced [see Eq.~\eqref{eq:rd}]. Also shown with
white lines are representative fluidlines, \ie radial free-fall orbits
with different fallback times, both during the outflow stage (positive
slope) and during the inflow stage (negative slope). Clearly, these
fluidlines intersect the trajectories of constant rest-mass density
shells since the fallback material is first decompressed in the outflow
stage and then re-compressed in the inflow part. Also apparent from
Fig.~\ref{fig:fig8} is that the vast majority of the ejecta has a
rest-mass density $\rho \lesssim 10^{-2}\,{\rm g/cm}^3$ as of $10\,{\rm
  s}$ after merger and that the bound matter with the largest but still
negative energy has the longest fallback time. Finally, shown with an
orange line in Fig.~\ref{fig:fig8} is the photosphere of the EM emission
model that we discuss in the next section.

%===================================================================
\subsection{Radiation from the fallback inflow}
\label{sec:ee}
%===================================================================

In order to determine the EM radiation arising from the fallback flow
discussed in the previous sections, we start by locating the photosphere
in the fallback flow. To this effect, we calculate the photospheric
radius (photosphere) $R_{\rm ph}(t)$, defined such that
\begin{equation}
\int_{r = R_{\rm ph}(t)}^\infty \rho(r, t) \kappa d r = 1\,,
\end{equation}
where $\kappa$ is the opacity of the material. Note that $R_{\rm ph}(t)$
marks the position of the photosphere \textit{within the fallback flow},
and is different from the that of the unbound and outward-moving material
that we discuss in Sec.~\ref{sec:discussion}.

\begin{figure}
  \center
  \includegraphics[width=0.15\linewidth, angle=-90]{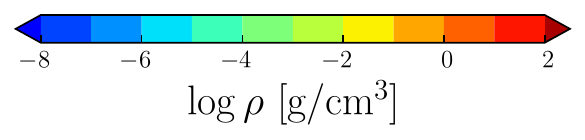}
  \includegraphics[width=\linewidth]{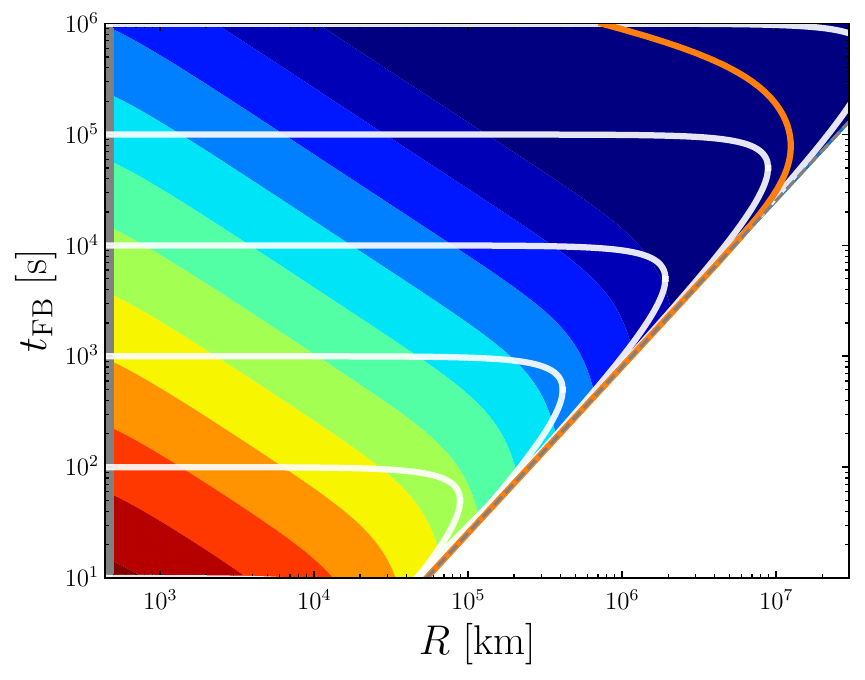}
  \caption{Spacetime distribution of the density in the fallback
    outflow-inflow as determined by our semi-analytical model
    (Sec.~\ref{sec:flow}). The white part in the lower-right corner is
    empty, because the bound material has not reached that altitude above
    the central object (Eq.~\eqref{eq:marg}). We present some fluid lines
    of the flow (white lines) and the photosphere used in the emission
    model (orange line). We show the surface of the detector sphere,
    where the fallback rate of $\dot{m}_{\rm FB} \propto t^{-5/3}$ is
    enforced (grey segment, Eq.~\eqref{eq:rd}) and the leading-edge of
    the outflow (grey dashed line). }
\label{fig:fig8}
\end{figure}

In the early stages of the post-merger that we have simulated, the
$r$-process nucleosynthesis is already well underway~\citep[see,
  \eg][]{Lippuner2015}, and the material is thus expected to already be
opaque~\citep[with $\kappa > 0.1\,{\rm cm}^2/{\rm
    g}$][]{Tanaka2020_opacity}, though exact opacity calculations at
these early times are lacking in the literature. Depending on the initial
electron fraction and thermodynamic properties of the material, the exact
composition and opacity will change. In addition, the composition of the
fallback column at different latitudes above the remnant will differ,
according to the differential irradiation by neutrinos, that our M1
scheme can capture, and we will consider below.

Assuming a low opacity of $\kappa = 0.5\,{\rm cm}^{2}/{\rm g}$, we report
with a solid orange line in Fig.~\ref{fig:fig8} the trajectory of the
photosphere in the spacetime diagram.  Clearly, over timescales $t_{\rm
  FB} \lesssim 10^4\,{\rm s}$, the photosphere coincides with the
leading edge of the bound ejecta, decoupling from it at later times, and
inverting its motion only at $t_{\rm FB} \lesssim \sim 10^5\,{\rm
  s}$. This behaviour is found even when considering a value of the
opacity as low as $\kappa = 0.1\,{\rm cm}^{2}/{\rm g}$, with the
photosphere moving inwards at times no earlier than $5 \times 10^4\,{\rm
  s}$ post-merger. As a result, when considering the larger opacity
reached through the $r$-process nucleosynthesis, we conclude that the
photosphere will always coincide with the edge of the ejecta, at least on
these timescales. Furthermore, the radiation emitted from the ejecta on
these timescales and coming from the leading-edge of the bound material,
can be modelled as having a black-body emission spectrum with a
time-varying temperature.

To capture the temperature evolution of this material in time and hence
account for $r$-process nucleosynthesis and expansion cooling, we use the
\texttt{SkyNet} nuclear-reaction network~\citep{Lippuner2017}. More
specifically, from our GRMHD simulations, we determine the temperature
$T$ and the electron fraction $Y_e$ of the material that is least bound,
\ie the material with largest $u_t$ among the bound matter crossing the
detector at $R_d$. This leading-edge material in the outflow is the
source of the radiation of interest, up to the recession of the
photosphere. Using our semi-analytical model, we extract the density
history at the leading edge $\rho(R_{\rm ph}(t), t)$ and determine the
initial nuclear statistical equilibrium at the initial density,
temperature and $Y_e$ using the nuclei libraries for strong reactions,
weak reactions, symmetric fission and spontaneous fission available in
\texttt{SkyNet}. Once this information is provided as initial conditions,
we use \texttt{SkyNet}'s reaction network to obtain the expected
temperature history $T_{\rm ph}(t)$ at the photosphere.

In principle, nucleosynthesis starts already from the unbinding of the
material from the merger remnant and is active already during the
propagation of matter up to detector $R_d$. This can be easily
appreciated after considering that the crossing time to the detector is
$\sim 1\,{\rm ms}$, which is larger than the neutron-capture timescale
\begin{equation}
  \tau_{n\,\rm cap} \gtrsim \frac{1}{v \, n_n \sigma_{n,\rm cap}} =
  10^{-8}\,{\rm s} \,,
\end{equation}
where we used a neutron-number density of $n_n = \rho / m_n =
10^{27}\,{\rm cm}^{-3}$ corresponding to a rest-mass density $\rho =
10^4\,{\rm g/cm^3}$ in the early stages of the flow expansion, a thermal
velocity $v = \sqrt{2 k_{\rm B} T / m_n}$ with temperature $T =
10^9\,{\rm K}$ and a cross section of neutron-capture onto protons of
$\sigma_{n,\rm cap} = 10^{-2}\,{\rm fm}^2$~\citep{Kopecky97}. In
practice, since the detector-crossing time is much smaller than the
overall timescale involved in the nucleosynthetic process\footnote{The
nuclear reaction networks show that the energy output from the
nucleosynthesis follows an evolution in time $\dot{e} \sim t^{-1.3}$ up
until tens of days, and then shut-off once the number of nuclei
participating in the reactions drops~\cite[see, \eg][]{Lippuner2017}.} it
is reasonable to ignore this initial contribution but more refined
calculations involving tracer particles~\citep[see, \eg][]{Bovard2016}
will help validate this assumption.

Knowing $R_{\rm ph}(t)$ and $T_{\rm ph}(t)$, we can determine the
black-body luminosity in any spectral band $\mathcal{B}$ from
\begin{equation}
L_{\rm FB}(t) = 8\pi R_{\rm ph}(t)^2 \int_{\nu \in \mathcal{B}}
B_{\nu}(T_{\rm ph}(t)) d \nu\,,
\end{equation}
where $\nu$ is the photon frequency and $B_{\nu}$ is the Planck function,
and we have clearly assumed the emission to be isotropic in view of the
spherical symmetry of the underlying model.

In our picture, the $r$-process heats up the photosphere, from which we
determine the radiation from the fallback flow. This approach neglects
the energy diffusion inside the ejecta and to judge whether this is
reasonable we recall that the photon-diffusion timescale inside the
infalling flow can be estimated as~\citep{Metzger2017}
\begin{equation}
\tau_{\rm diff} = \kappa {R_{\rm ph}^2 \rho_{\rm ph}}\,,
\end{equation}
with $\rho_{\rm ph}$ the rest-mass density at the photosphere. Using an
opacity of $\kappa = 1\,{\rm cm}^2{\rm /g}$ and data read off
Fig.~\ref{fig:fig8}, this timescale decreases from $\tau_{\rm diff}
\sim\,10^7\,{\rm s}$ at $t-t_{\rm mer} = 10\,{\rm s}$ post-merger to
$\tau_{\rm diff} \sim\,10^4\,{\rm s}$ at $t-t_{\rm mer} = 10^4\,{\rm s}$;
obviously, the photon diffusion timescale would be even higher with a
larger opacity. Overall, these considerations show that $\tau_{\rm diff}$
exceeds the timescales over which we study the radiation (\ie $\lesssim
10^4\,{\rm s}$), such that in our simplified model we can neglect the
thermalisation in the outflow, and consider the temperature history from
the nuclear reaction network output as the photosphere temperature.

\begin{figure*}
  \center
  \includegraphics[width=0.48\textwidth]{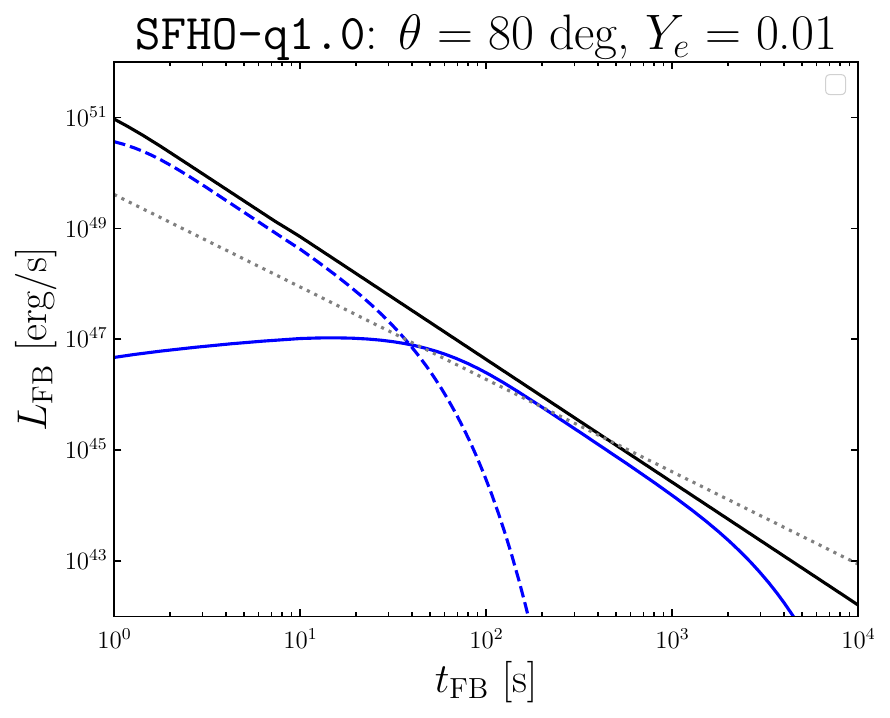}
  \includegraphics[width=0.48\textwidth]{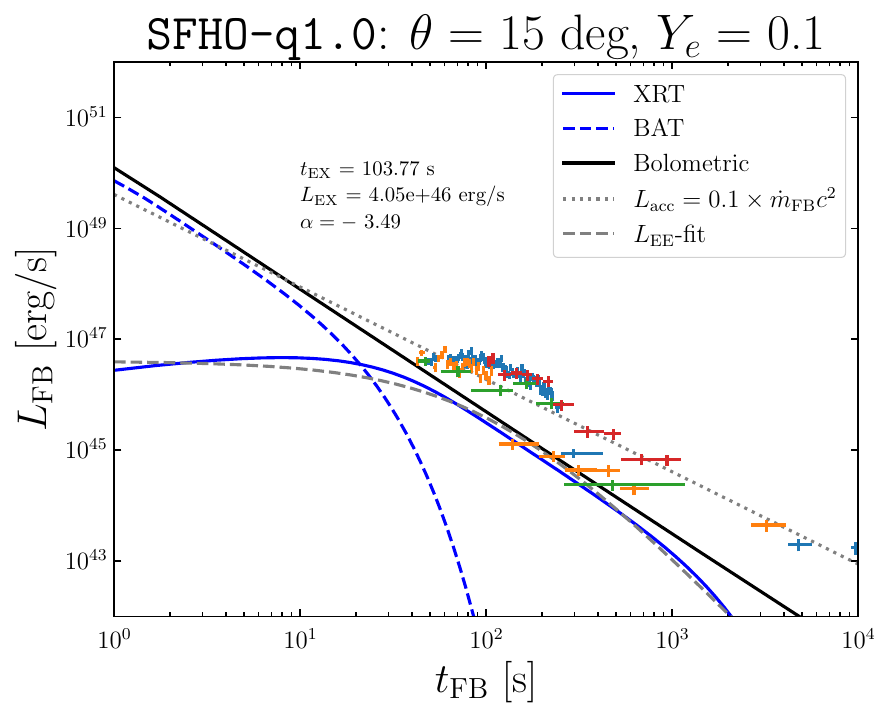}
  \includegraphics[width=0.48\textwidth]{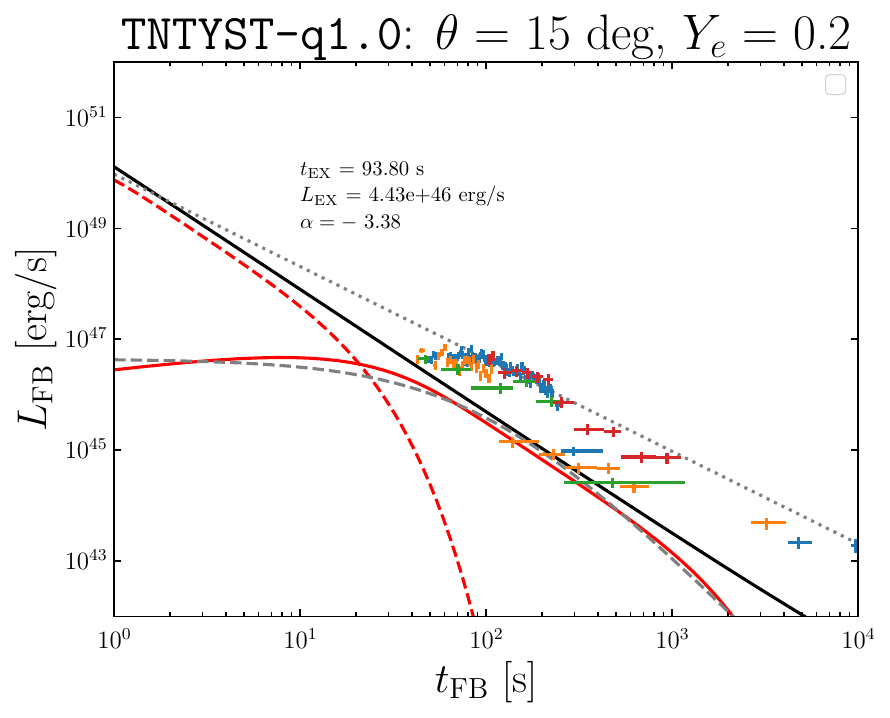}
  \includegraphics[width=0.48\textwidth]{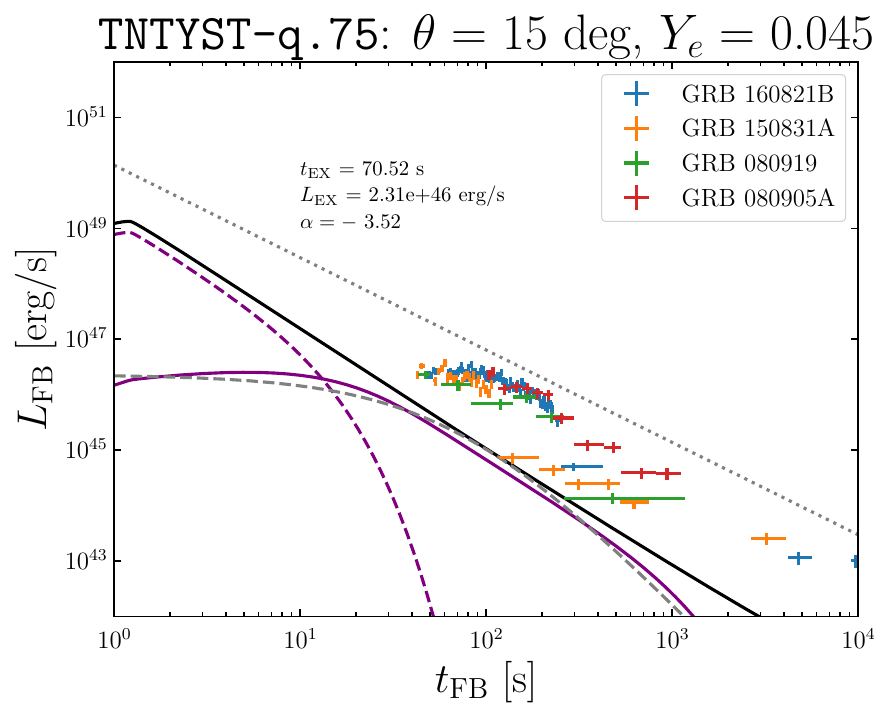}
  \caption{Radiation arising from the fallback flow as per our model of
    black-body emission from the photosphere heated up by $r$-process
    nucleosynthesis (Sec.~\ref{sec:ee}). In each panel, the temperature
    evolution was determined from the $r$-process heating starting from a
    $Y_e$ value measured at different latitudes in the fallback flow, as
    indicated above the panels. We present the luminosity in the
    \textit{Swift}/BAT and \textit{Swift}/XRT bands (coloured solid and
    dashed lines, respectively) and the bolometric luminosity (black
    solid line). We also present the luminosity derived from the
    accretion rate (grey dotted line). In the panels corresponding to
    near-polar views ($\theta = 15\,{\rm deg}$), we present the best-fit
    model of a flat-power-law light-curve to the XRT luminosity (grey
    dashed line, Eq.~\eqref{eq:leee}). For these panels, the best-fit
    parameters are reported in the figures. Finally, we show the XRT data
    for some GRBs with extended emission in the source frame, arbitrarily
    scaled in flux to compare with our model (coloured
    points). }
\label{fig:fig9}
\end{figure*}

In Fig.~\ref{fig:fig9}, we plot the bolometric luminosity and the
luminosity integrated over the \textit{Swift}/XRT and \textit{Swift}/BAT
bands, namely, [0.3--10]\,keV and [15--350]\,keV,
respectively~\citep{Gehrels2004, Barthelmy2005b}. All the curves in these
plots correspond to times before the recession of the photosphere. In the
different panels, we consider the fallback inflow deduced from our four
GRMHD simulations, and with fallback material picked from different
latitudes $\theta$ in the inflow, as marked in the panels. The initial
value of the electron $Y_e$ of that material, also marked in each plot,
is determined from our M1 scheme. The dotted line marks the bolometric
luminosity from the fallback accretion that one would deduce from a naive
conversion from the fallback rate $L_{\rm acc} = \eta \dot{m}_{\rm FB}
c^2$, with an efficiency of $\eta = 10\%$ and $\dot{m}_{\rm FB}$ as in
Fig.~\ref{fig:fig6} (see discussion at the end of Sec.~\ref{sec:eos}).

When inspecting the various panels is Fig.~\ref{fig:fig9} a number of
considerations follow. First, the duration and luminosity of the fallback
radiation are significant and comparable with the observed extended
emission in short GRBs (see Secs.~\ref{sec:intro} and
\ref{sec:discussion}). Second, through a non-trivial interplay between
the photospheric radius and the temperature evolution, the bolometric
luminosity displays a power-law decay with nearly constant slope $\alpha
\sim -8/3$, which is steeper than the decay of $\dot{m}_{\rm FB} \propto
t^{-5/3}$. Inspecting the temperature evolution from the nuclear-reaction
network reveals that the temperature evolves as $T \propto t^{-1}$, and a
heating rate that also has a power-law behaviour as $\epsilon_{\rm
  nuc.}(t) \propto t^{-3/2}$, in agreement with other studies of
$r$-process nucleosynthesis \citep{Lippuner2015, Wanajo2014}.

Because of the temperature decrease with time, the radiation is first
mainly output in the gamma-ray bands before dominating the X-ray band
when the Planck function's maximum crosses the band. At very early times,
\ie for $t < 1\,{\rm s}$, the \textit{Swift}/BAT band is below the
black-body peak temperature $T \gtrsim 100\,{\rm KeV}$, such that the
Rayleigh regime applies with $L_{\gamma} \sim R^2 T \propto
t^{1/3}$. This near-constant segment of the light-curve is outside of the
plot in Fig.~\ref{fig:fig9}, as it is not expected to be visible due to
the prompt emission outshining this component. However, an analogous
segment is seen in the X-rays, which lasts until tens of seconds, with a
near-flat X-ray light-curve. Once the black-body temperature reaches the
\textit{Swift}/BAT band, at $t\sim 1\,{\rm s}$, the gamma-ray luminosity
follows the bolometric behaviour with $L_{\gamma}\propto^{-8/3}$. This
emission lasts until the black-body peak frequency leaves the
\textit{Swift}/BAT band, at nearly one hundred seconds. After that, an
exponential cutoff is seen in the gamma-rays, leading to the $t^{-8/3}$
segment in the X-rays up to a thousand seconds, followed by the
exponential cutoff. In general, the sequence
``$t^{1/3}$''-plus-``$t^{-8/3}$''-plus-``exponential cutoff'' is present
in all bands, however the earlier segment in the gamma-rays and the later
segment in the X-ray are expected to be outshined by the prompt and the
forward-shock afterglow respectively.

\begin{figure*}
  \center
  \includegraphics[width=0.65\textwidth]{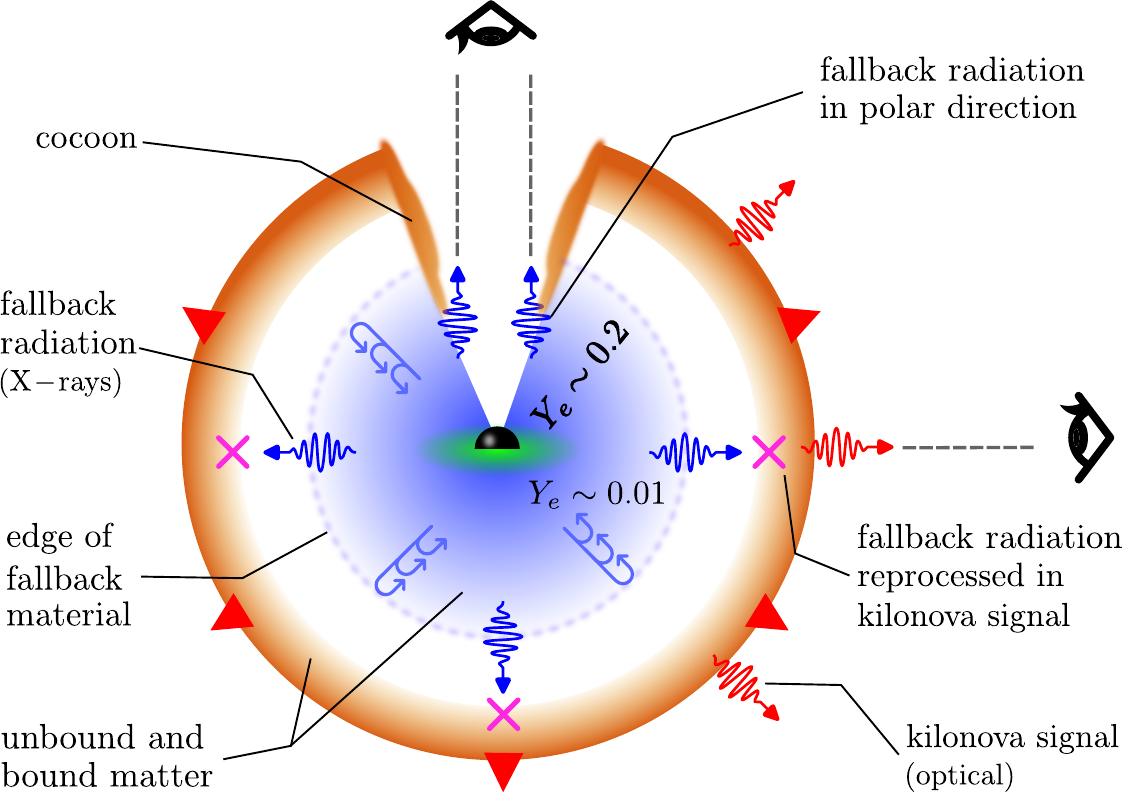}
  \caption{A schematic view of our model for the EM signature of the
    fallback flow. The fallback outflow-inflow (blue) is ejected
    underneath the unbound material (purple), where the kilonova signal
    is emitted (red photons). The fallback material emits X-rays (blue
    photons), that can either be absorbed and reprocessed in the kilonova
    ejecta (equatorial direction), or reach a distant observer by passing
    through the free way opened by the passing of the jet launched
    post-merger (polar direction). In this case, the fallback radiation
    appears as high-energy extended emission.}
\label{fig:cartoon}
\end{figure*}

Since the photospheric radius at early times is determined only by the
orbit with $u_t = 0$ and is therefore essentially the same for all
binaries considered, differences in luminosity between binaries and
different latitudes in Fig.~\ref{fig:fig9} those binaries are simply due
to the initial thermodynamic conditions and $Y_e$. For example, when
comparing the luminosity curves at different latitudes within a same
binary (\eg the two top panels in Fig.~\ref{fig:fig9}) it is possible to
note the effect of a higher initial $Y_e$: the binary with a lower $Y_e$
experiences a stronger heating rate and a brighter emission. On the other
hand, when comparing binaries with similar initial $Y_e$ and different
total fallback masses (\eg the two right panels in Fig.~\ref{fig:fig9}),
it is possible to appreciate the impact of the history of the rest-mass
density: the overall luminosity is higher for the higher-mass system.

After the photosphere recedes into the ejecta, our simple emission model
no longer applies, as the whole ejecta then shines. The mass of emitting
material then follows $m_{\rm shine} \propto t^{-2/3}$ and, applying a
similar temperature evolution $T\propto t^{-1}$ to the whole ejecta, one
can expect a leveling-out of the luminosity with a shallower decay than
in Fig.~\ref{fig:fig9}. This segment, however, is likely to be
unobservable because of the forward-shock afterglow.

%-------------------------------------------------------------------
%-------------------------------------------------------------------
\section{A new high-energy emission component from BNS mergers}
\label{sec:discussion}
%-------------------------------------------------------------------
%-------------------------------------------------------------------

The considerations made so far were rooted on accurate and reasonably
realistic state-of-the-art simulations of BNS mergers. In what follows we
take the lessons learned from this analysis to develop a model for a new,
high-energy extended-emission component from BNS mergers.

To illustrate simply our general extended-emission model, we make use of
the cartoon shown in Fig.~\ref{fig:cartoon} and recall that matter is
ejected during the merger and post-merger phases in terms of material
that is mostly unbound at first and is instead mostly bound at later
times (see Fig.~\ref{fig:fig3}). The expanding unbound material,
represented in orange in the cartoon, is thus spatially separated from
the bound material that will fallback with highly eccentric orbits, in a
quasi-spherical fashion (see Sec.~\ref{sec:slope}), and is represented in
blue in the cartoon. The unbound material will produce the kilonova
optical transient and it is a (very) optically thick layer of matter
between the fallback accretion and the external observers. In the case in
which a jet is launched by the remnant and successfully breaks out from
the merger ejecta, it will drive a cavity in the polar direction. The
walls of this cavity are made up of the so-called
cocoon~\citep{Bromberg2011, Matsumoto2019}, that is, a relatively
low-density region where the jet has deposited energy in its interaction
with the ejecta.

As a result, and as sketched in the cartoon, the observer's ability of
receiving the EM radiation from the fallback inflow will depend
significantly on the inclination angle between the observe and the
direction of propagation of the jet. Setting such an angle to be the
latitude of the ejecta, it is clear from the cartoon that at high
latitudes, \ie along directions that are equatorial or nearly equatorial,
this radiation is trapped by the kilonova material and hence it will not
reach the observer. Rather, it will be absorbed and reprocessed,
energising the material, with consequences on the kilonova signal (see
discussion at the end of this section). On the other hand, at low
latitudes, \ie along the polar directions, the essentially baryon-free
free region opened by the jet or the lower optical thickness of the
cocoon can allow the fallback radiation to escape and reach a distant
observer. Indeed, we conjecture that, as the polar direction and cocoon
rarefy due to expansion, the fallback radiation for $t_{\rm FB} \gtrsim
10\,{\rm s}$ will emerge, thus making up for a hitherto unconsidered
emission component in BNS mergers.

The separatrix between the trapped and the escaping fallback radiation
will depend of a number of factors, such as the physical separation
between the bound and unbound flows, the opening of the jet cavity, the
hydrodynamical properties of the ejecta and its degree of anisotropy, as
well as the size and density of the cocoon. Overall, simulations show
that the angular size of the cocoon is small~\citep{Hamidani2021}, such
that the vast majority of the ejecta is unperturbed by the passing of the
jet and should follow the description of Sec.~\ref{sec:3}, especially for
very luminous jets that successfully break-out from the
ejecta~\citep{Duffell2018, Mpisketzis2024}.

Hence, it is reasonable to expect that along lines-of-sight that are
essentially polar, as those expected to be involved in the observations
of short GRBs at high redshifts~\citep{Beniamini2019}, a high-energy
emission component with features as displayed in Fig.~\ref{fig:fig9} will
emerge sometime after the detection of the gamma-ray component and of the
forward-shock afterglow. The overall luminosity will clearly depend on
the various factors above-mentioned as these will determine how much of
the fallback radiation is made actually visible. Furthermore, while the
light-curves at different latitudes and different composition evolution
will blend, the basic feature of the radiation discussed in
Sec.~\ref{sec:ee} will represent the basic features, namely, gamma-ray
emission extending for up to a hundred seconds ending in a sharp cut-off
followed by a X-ray segment lasting for up to hundreds of
seconds. Clearly, a new EM-emission component with these properties
represents a very good candidate for extended emission in short GRBs.

To further corroborate this suggestion, we recall that an essential
feature of extended emission in short GRBs is the relatively softer
spectrum with respect to the parent prompt emission~\citep[see,
  \eg][]{Barthelmy2005, Gehrels2006_ee, Kaneko2015_ee}. Indeed, as shown
in Fig.~\ref{fig:fig9}, in our model, the extended-emission component in
the gamma-ray band is dominated by the $t^{-8/3}$ segment of the
gamma-ray light-curve. This corresponds to the crossing of the black-body
peak in the gamma-ray band, during which the hardness ratio quickly
decreases. Moreover, the ratio of the extended-emission energy $E_{\rm
  EE}$ to the and prompt emission energy $E_{\rm PE}$ in gamma-rays
varies in a broad range, from $E_{\rm EE} / E_{\rm prompt} \lesssim 0.1$
to $E_{\rm EE} / E_{\rm prompt} \gtrsim
40$~\citep[\eg][]{Bostanci2013_ee, Kagawa2019_ee}. Beyond the uncertain
geometrical factor that will condition the visibility of the fallback
radiation in our scenario, we note that our model attributes these
emissions to two distinct flows in BNS mergers, with energy output that
are somewhat unrelated. Also, as we pointed out in Fig.~\ref{fig:fig9},
the initial electron fraction of the bound material determines the
temperature evolution and thus the time of the cutoff in the gamma-ray
extended-emission luminosity. As a result, the duration of the
extended-emission episode is in our picture is weakly correlated with the
temporal features of the prompt-emission, which remains to be explored in
extended-emission catalogues. When considering instead the X-ray
component of the extended emission, we note that the exponential cutoff,
generally attributed to a black-hole spin-down in second-jet
scenarios~\citep{Kagawa2015_xray}, in our model is due to the crossing of
the black-body peak frequency in the X-ray band. Again, this will occur
sooner or later according to the heating rate in the bound material. As a
result, we expect no strong correlation with the temporal properties of
the prompt emission.

Finally, when contrasting our fallback model with the extended-emission
models involving a magnetar emission~\citep[see, \eg][]{Metzger2008,
  Bucciantini2012_ee, Rezzolla2014b, Ciolfi2014}, the observation of the
X-ray radiation from the magnetar is also subject to the opening of a
cavity in the kilonova ejecta, or, in any case, to a relatively small
optical thickness in the polar direction. Since the drilling of such a
cavity is subject to launching a powerful and ultrarelativistic jet that
breaks out, the magnetar picture must reconcile a (very) long-lived
neutron star with the launching of a very luminous jet~\citep[see,
  \eg][for a discussion of this possibility]{Ciolfi2017, Ciolfi2020,
  Moesta2020, Combi2023}.

Obviously, any theoretical model is only as good as it is able to
reproduce naturally the actual observations. Hence, as a first comparison
with the observations, we follow~\citet{Kisaka2017} in fitting the
\textit{Swift}/XRT photometry expected from our model with a plateau
followed by a power-law decay
\begin{equation}
L_{\rm EE}(t) = L_{\rm EX} \left( 1 + \frac{t}{t_{\rm
    EX}}\right)^{\alpha} \,,
\label{eq:leee}
\end{equation}
defining the luminosity and duration of the extended emission episode in
the X-rays.  Here, ${L}_{\rm EX}$ and ${t}_{\rm EX}$ are fitting
parameters representing the luminosity and duration of the extended
emission episode. However, contrarily to~\citet{Kisaka2017}, we do not
set the temporal index $\alpha$ and leave it as a free parameter. Indeed,
Fig.~\ref{fig:fig9} suggests that $\alpha$ is steeper than $-5/3$ (the
fallback accretion rate index) and shallower than the $-40/9$ predicted
by the secondary-jet model~\citep{Kisaka2015_jets}. The best-fit model to
$L_{\rm EE}$ is shown for all the binaries in Fig.~\ref{fig:fig9} with
the grey dashed line (consistent with our discussion on the conditions
for the observability of the fallback emission, we only consider models
with polar views in this comparison with GRB data).

Indeed, we find that the functional form for $L_{\rm EE}$ provides a good
fit to our predictions, allowing a comparison with the catalogue
from~\citet{Kisaka2017} and where the best-fit parameters are reported in
Fig.~\ref{fig:fig9}. In this catalogue, the events that have a secure
redshift present a clustering both in terms of their duration and their
luminosity of the extended emission episodes\citep[see Fig.~2 of][panels
  C and D, blue histogram]{Kisaka2017}. We find that the best-fit $t_{\rm
  EX}$ and $L_{\rm EX}$ of our cases presented in Fig.~\ref{fig:fig9}
fall within these clusters of the catalogue. Concerning the events without
known redshift in the catalogue (their panels A and B), the best-fit
parameters are also consistent, though the clustering in the catalogue is
less clear in this case. In any case, we find that the temporal slope
that best fit out light-curves is shallower than $-40/9 \sim -4.4$, with
$\alpha \sim -3.5$ in all cases, consistently with the temporal decay of
the bolometric luminosity discussed previously.

To further ground our fallback-emission model in the observations of
extended emission from short GRBs, we report on the light-curve plots in
Fig.~\ref{fig:fig9} some data from the extended emission in short GRBs
observed by \textit{Swift}. The dataset was selected among the events
with extended-emission components from~\citet{Kisaka2017} and chosen
since they present a shallower decrease than the $-40/9$ employed in
their fits. The sources in question are: GRBs\,080905A, 080919, 150831A
and 160821B, and the corresponding data was retrieved from the XRT
database~\citep{Evans2007_xrt, Evans2009_xrt}. The position in time is
obtained using the redshifts reported by~\citet{Kisaka2017} and after
determining the times of observation in the source-frame as $t_{\rm RF} =
t_{\rm obs}/(1 + z)$, with $t_{\rm RF}$ and $t_{\rm obs}$ the time
coordinate in the source frame, which is redshifted by $z$, and in the
observer frame on Earth, respectively. On the other hand, the XRT fluxes
for the different GRBs were arbitrarily rescaled, each with a different
factor, to compare the data with our light-curves. This allows us to
focus on the shape of the light-curves, regardless of their overall flux.

Overall, Fig.~\ref{fig:fig9} shows that these shallow-decaying
extended-emission episodes are perfectly consistent with the new
extended-emission model presented here. Furthermore, a closer look at the
light-curves from~\citet{Kisaka2017} reveals that the extended-emission
episodes actually exhibit a larger variety of decay rates than suggested
by the single and ``universal'' index of $-40/9$ suggested
by~\citet{Kisaka2017}. Indeed, the power-law decay ranges from very
shallow values (as measured for GRBs\,060801, 061006, 100724A, 090426),
to shallow (as in the selection of GRBs mentioned above), and up to steep
(as it is the case for GRBs\,071227, 081023, 160821B). These
considerations suggest that it is reasonable to consider a diversity of
physical mechanisms with varying decay behaviours can be invoked to
describe the extended-emission phenomenon, including a fallback accretion
scenario or a secondary black-hole jet. We also note that the
non-uniqueness of the physical origin of extended-emission X-ray
components would also provide a natural explanation for the apparent
bimodality in the duration and luminosity of these phenomena
\citep{Kisaka2017}.

As anticipated in Sec.~\ref{sec:ee}, in addition to the intrinsic
evolution of the radiation from the fallback material, the fact that the
photosphere eventually starts to recede back to the central object
suggests that, at least in principle, it will be eventually possible to
``reveal'' the black hole and eventually another regime of optically-thin
spherical accretion. However, whether this is possible in practice,
depends also by the forward-shock afterglow, which could easily outshine
the very late-time accretion luminosity and that should be $L <
10^{42}\,{\rm erg/s}$ in the X-ray band.

We conclude this section with two final but important remarks. First, in
contrast to the magnetar and secondary-jet models~\citep[see,
  \eg][]{Barkov2011_ee, Kisaka2015_jets}, our picture for extended
emission decouples the duration of the extended emission from the
timescales of the small-scale physics near the remnant. Indeed, the
magnetar picture requires the magnetar to survive and shine for as long
as the extended emission lasts, that is, up to several hundreds of
seconds. Similarly, the secondary-jet picture requires the second
jet-launching episode to last for this duration. In our case, it is only
a single and very short-lived process (the ejection of bound material
over $t_{\rm ej} \lesssim 100\,{\rm ms}$) that results in a long-lived
emission episode. The characteristics of this emission are set only from
the very early post-merger dynamics and, as illustrated in our sample of
binaries, it appears to be rather robust. In this respect, while our
model of extended emission from fallback material may not explain all of
the observations, it is clear that it represents a natural, high-energy
component to be expected in the emission from BNS mergers and hence short
GRBs.

Second, we note that, depending on the radiative efficiency of the
fallback flow and on the efficiency of reprocessing of this radiation
into the kilonova emission, the latter can be affected and potentially in
a significant manner. In particular, it is reasonable to expect that the
fallback radiation can change the kilonova luminosity by roughly the
ratio of fallback mass to unbound mass, \ie $\eta_{\rm FB} := m_{\rm FB}
/ m_{\rm KN}$. As reported in Tab.~\ref{tab:1}, this ratio can actually
reach up to $\sim 50-60\,\%$ and is systematically larger for asymmetric
binaries or binaries with long-lived remnant central objects. Clearly,
such a powerful emission has the potential of impacting the observed
kilonova signal and the inference of the properties of the emitting
system. A particular example of this impact is given by the inference of
the lifetime of the remnant central object in GW170817. \citet{Gill2019}
used a series of GRMHD simulations of the various magnetic-field- and
neutrino-driven ejection channels to determine how long the remnant of
the GW170817 event should have lasted to eject the mass of
lanthanide-poor material inferred from the observations of the kilonova
signal, that is, the ``blue'' component.  The analysis carried out
by~\citet{Gill2019} lead to the conclusion that the remnant should have
survived for a rather long time, \ie $t_{\rm coll} \simeq 1\,{\rm s}$,
before collapsing to a black hole. Although similar estimates have been
confirmed also by other authors~\cite[see, \eg][]{Murguia-Berthier2020},
such long-lived remnant do require a rather large lanthanide-poor ejected
mass. On the other hand, a smaller blue-component mass would actually be
needed to be ejected if the kilonova signal was modeled to contain an
additional source of energy coming from the reprocessing of the radiation
from the fallback material. Overall, this example points out that the
modelling of kilonova signals (and therefore the inference of ejecta
outflows from kilonova signals), which already suffers from a number of
uncertainties~\citep[see, \eg][]{Barnes2021_kn, Bulla2023}, is rendered
even more difficult by the effect of energy injection from the fallback
material, which should be considered in the future.

%-------------------------------------------------------------------
%-------------------------------------------------------------------
\section{Conclusion}
\label{sec:conclusion}
%-------------------------------------------------------------------
%-------------------------------------------------------------------

The vast majority of the works that has investigated the inspiral, merger
and post-merger of binary systems of neutron stars has concentrated on
the matter that is ejected from the system as gravitationally
\textit{unbound} and is therefore responsible for the generation of the
kilonova signal and for the nucleosynthesis of heavy elements. The matter
that is instead gravitationally \textit{bound} has not received much
attention, as it has been normally assumed to be either as too small to
be astrophysically relevant or not to produce a contribution to the total
EM signal from merging binaries.

By making use of accurate and state-of-the-art GRMHD simulations
including proper neutrino transport, we have shown that both of these
assumptions are actually incorrect. More specifically, by studying the
inspiral and merger of four different binary systems spanning different
mass ratios and EOSs, we have found that the amount of bound matter is
actually quite substantial, being almost $50\%$ of the unbound matter and
reaching a total of $\gtrsim 10^{-3}\,M_\odot$. Furthermore, the
accretion rate follows a universal power-law in time with slope $\simeq
t^{-5/3}$, which is independent of the EOS, the properties of the binary
and the fate of the remnant.

Interestingly, the timescale of the fallback and the corresponding
accretion luminosity all are in good agreement with the long-term
emission observed in short GRBs and that is normally referred to as the
``extended emission''. The origin of such an emission is still largely
unclear and represents one of the most challenging aspects of the
modelling of the EM emission from short GRBs. Using the information from
the simulations, and employing a semi-analytical treatment of the
fallback dynamics, we have been able to study in detail the fallback
accretion and the radiation arising from such inflow and to explore the
possibility that extended emission can be attributed to this fallback
material. More specifically, using a simple electromagnetic emission
model that is based on the thermodynamical state of the fallback material
heated by r-process nucleosynthesis we have shown that the fallback
material can shine in the gamma- and X-rays with luminosities $\gtrsim
\,10^{48}\,{\rm erg/s}$ for hundreds of seconds, thus making it a good
and natural candidate to explain the extended emission in short
GRBs. Furthermore, such emission reproduces well and rather naturally
some of the phenomenological traits of the extended emission, such as its
softer spectra with respect to the prompt emission and the presence of
exponential cutoffs of the luminosity in time.

On lines-of-sight aligned with the remnant polar axis, and thus looking
through the funnel drilled by the relativistic jet, this emission
component is a good candidate for the extended-emission of short
gamma-ray bursts. On the other hand, along lines-of-sight that
significantly misaligned with the polar axis, the fallback radiation
represents another source of energy injected into the ejecta. As a
result, for these inclinations, the energy injection from the fallback
material should be considered as an additional source of uncertainty in
the modelling of kilonova signals, and therefore the inference of ejecta
outflows from kilonova signals.

Finally, explaining the extended emission in terms of the luminosity
produced by the fallback accretion of bound matter has an important
advantage over alternative explanations. Indeed, the magnetar picture
requires the magnetar to survive and emit radiation for as long as the
extended emission lasts, that is, up to several hundreds of
seconds. Similar considerations apply also when considering the extended
emission being produced by the secondary jet. In the interpretation
proposed here, and thanks to the power-law $t^{-5/3}$ decay of the
accretion rate, we can invoke a single ejection episode over a
comparatively short window in time of $\mathcal{O}(100)\,{\rm ms}$ to
obtain an emission that can be sustained for up to $10^3-10^4\,{\rm s}$.

In summary, our results clearly highlight that fallback flows onto merger
remnants cannot be neglect and represent a very promising and largely
unexplored avenue to explain the complex phenomenology of GRBs. This
exciting prospect calls for a more detailed analysis of the scenario
proposed here and for a number of improvements in our model. These
include: a more accurate description of the fallback flow to account for
deviations from spherical symmetry \new{and the corresponding
  back-reaction of the emitted radiation onto the accreting matter,} a
more refined model of the emitted radiation \new{and of its anisotropic
  interaction with the material in the cocoon}, a detailed study of the
offset in time between the merger time and the start time of the
extended emission, and the inclusion of the interaction between the
accreting and the ejected material when the ejection lasts over
timescales much larger than $100\,{\rm ms}$. We plan to address these
points in future studies.

%-------------------------------------------------------------------
%-------------------------------------------------------------------
\section*{Acknowledgments}
%-------------------------------------------------------------------
%-------------------------------------------------------------------

It is a pleasure to thank Jonathan Granot for useful input and comments,
and a careful reading of the manuscript. This research is supported by
the European Research Council Advanced Grant ``JETSET: Launching,
propagation and emission of relativistic jets from binary mergers and
across mass scales'' (grant no. 884631). LR acknowledges the Walter
Greiner Gesellschaft zur F\"orderung der physikalischen
Grundlagenforschung e.V. through the Carl W. Fueck Laureatus Chair

%-------------------------------------------------------------------
%-------------------------------------------------------------------
\section*{Data availability}
%-------------------------------------------------------------------
%-------------------------------------------------------------------

The data and software underlying this article will be shared on
reasonable request to the corresponding author.

\section*{Appendix}
\label{sec:Appendix}

%-------------------------------------------------------------------
%-------------------------------------------------------------------
\subsection{Going beyond the simulations runtime}
\label{sec:B}
%-------------------------------------------------------------------
%-------------------------------------------------------------------

The analysis presented here is inevitably limited by the duration of our
simulations. While the two SFHO binaries have only a minute mass ejection
by the time our simulations end at $t\simeq 50\,{\rm ms}$,
(Fig.~\ref{fig:fig3}), this is not the case for the TNTYST binaries,
which do not lead to a black-hole, and where the ejections is nonzero at
the time the simulations are terminated. Although Fig.~\ref{fig:fig7}
clearly suggests that the late ejecta has very short fallback timescales,
we cannot conclude that ejection on longer timescales than the ones we
capture in our simulations would not contribute to longer fallback
times. Indeed, if this late-time ejection had longer fallback times, the
EM signature would likely be different.

\begin{figure}
  \center
  \includegraphics[width=0.48\textwidth]{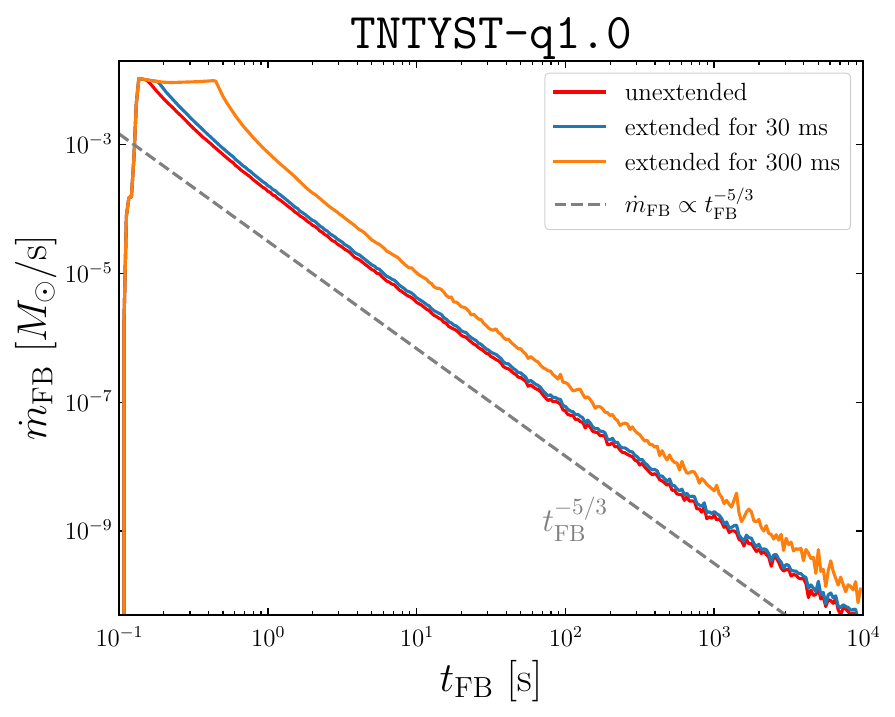}
  \caption{Fallback accretion luminosity obtained by extending the
    ejection history of the \texttt{TNTYST-q1.0} system. The solid lines
    represent the fallback accretion luminosity for different lengths of
    the ejection extension. The ejection segment that is reproduced to
    make up the extension is presented in Fig.~\ref{fig:fig7} (grey
    shaded region).}
\label{fig:figB}
\end{figure}

To explore this possibility, we virtually extend the duration of our
simulations by considering ejection of matter also at times past the end
of our simulations. We do this by replicating the ejection of matter from
the \texttt{TNTYST-q1.0} simulation and relative to a data segment
spanning a window in time of $10\,{\rm ms}$. This segment is shown with a
grey-shaded area in the bottom-left panel of Fig.~\ref{fig:fig7} and
refers to an ejection of matter with almost steady-state fallback
properties (see also Fig.~\ref{fig:fig6}). We then replicate this data
segment a number of times, \ie $3, 30$ or $300$ times, hence mimicking a
simulation that would run for $\simeq 0.053$\footnote{The actual
simulation lasts for about $50\,{\rm ms}$.}, $0.3$, and $3.0$ seconds,
respectively. We then determine the corresponding fallback mass accretion
rate and show the results in Fig.~\ref{fig:figB}, where different
solid-coloured lines refer to different ejection times. By contrasting
Figs.~\ref{fig:fig6} and~\ref{fig:figB}, it is clear that the ``extra''
ejection mimicked here has the effect of introducing a nearly-flat
accretion segment that lasts for the duration of the extra ejection
before connecting to the $t^{-5/3}$ fall-off for the subsequent times and
as described in Sec.~\ref{sec:slope}. This mass-accretion stage would
lead to a similar phase of constant observed luminosity. Thus, the
observational signatures of the bound material are not tied to the
timescales of the extended emission, but can be prolonged to longer
emission components, such as some of the early plateaus in X-ray
afterglows of GRBs~\citep{Nousek2006}.

As a word of caution we should remark that when modelling ejection
timescales that are much larger than the one considered in the
simulations (\ie $\mathcal{O}(100\,{\rm ms})$), and indeed of the order
of seconds as considered in Fig.~\ref{fig:figB}, then it would be
necessary to account also for the interaction of the late-time ejected
matter with the infalling matter. The two components could collide
leading to a reverse shock and potentially a new component of the
fallback luminosity. The much richer picture that could follow from this
additional component is worth exploring in future works.

%-------------------------------------------------------------------
%-------------------------------------------------------------------
\subsection{Newtonian and general-relativistic fallback times}
\label{sec:A}
%-------------------------------------------------------------------
%-------------------------------------------------------------------

We next compare our method of determining the fallback time of bound
material using a Newtonian description for the fallback orbits ($t_{\rm
  FB, N}$, see Eqs.~\eqref{eq:epsilon}--\eqref{eq:orb}) with a
relativistic treatment using the integration of timelike geodesics
representing the orbits of massive test particles in a Kerr spacetime.

Taking as a reference the \texttt{SFHO-q1.0} binary, we perform a
quasi-local measurement~\citep{Thornburg95, Dreyer02a} of the mass
$M_{\rm BH}$ and dimensionless spin $\chi_{\rm BH}$ of the black hole
that forms as a result of the collapse of the HMNS. Since the mass of the
torus around the black hole is a few percent at most of the mass of the
black hole~\cite{Rezzolla:2010}, we can ignore it and hence use $M_{\rm
  BH}$ and $\chi_{\rm BH}$ to build a Kerr spacetime that we cover with
Cartesian Kerr-Schild coordinates having $\boldsymbol{g}_{\rm KS}$ as
metric tensor~\citep[see, \eg][]{Kerr:2009}. We then collect fluid
elements on our detector sphere at $R_d = 300\,M_\odot$ and read off
their energy $u_t$ and the local spatial components $u_i$ which we then
rescale for the different coordinate system and so as to guarantee that
the four-velocity is a unit timelike vector, \ie that
$(g_{\alpha\beta})_{\rm KS} u^\alpha u^\beta = -1$. Using these initial
conditions, we integrate the corresponding geodesic equations in
Kerr-Schild coordinates until the material reaches the detector surface
again, which we set at $R_{\rm BL} = 300\,M_\odot$, where $R_{\rm BL}$ is
the radial Boyer-Lindquist coordinate \citep{Boyer1967}. In this way, we
can determine the corresponding fallback time $t_{\rm FB}$ as the time
coordinate when this terminal condition is met. We apply this procedure
to the $\sim\,70,000$ fluid elements crossing our 2-sphere over a single
timelevel and find they span the range in specific energy $\epsilon \in
[- 1 \times 10^{-2}, -3 \times 10^{-5}]$, which is representative of the
material relevant for our study.

\begin{figure}
  \center
  \includegraphics[width=0.48\textwidth]{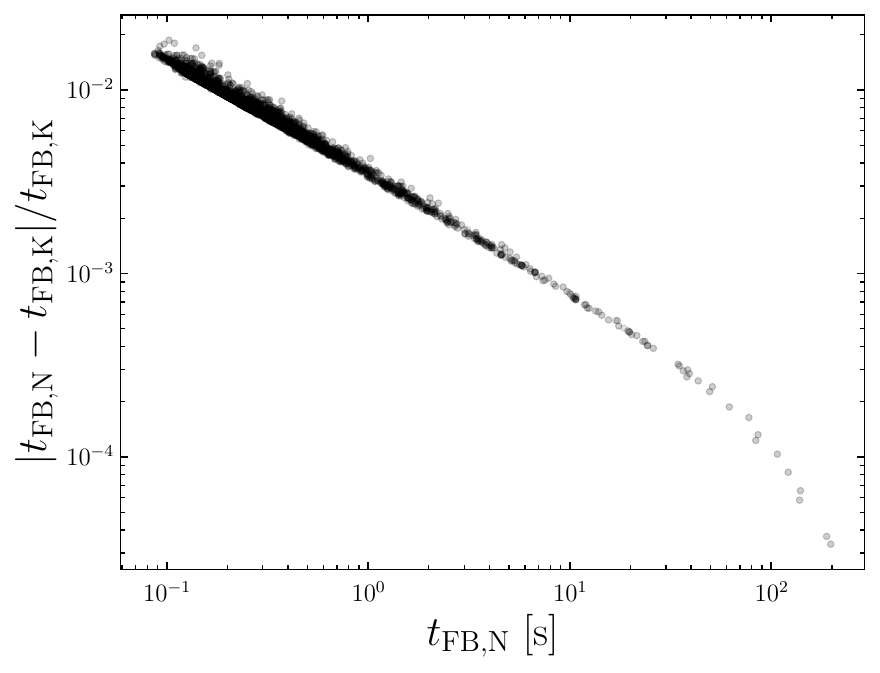}
  \caption{Comparison between the fallback time estimated by a Newtonian
    description of the fallback orbits ($t_{\rm FB, N}$) with that
    estimated by a relativistic integration of the orbits in a Kerr
    spacetime ($t_{\rm FB, K}$). This comparison serves to justify our
    use of the Newtonian estimate throughout our study.}
\label{fig:figA}
\end{figure}

Figure~\ref{fig:figA} reports the relative difference between the
general-relativistic and Newtonian fallback times for these particles,
\ie $t_{\rm FB, N}$ and $t_{\rm FB, N}$, respectively. Naturally, the
difference is smaller for longer orbit times, that spend more time far
away from the black hole, as differences in the gravitational fields are
much smaller far from the black hole. Overall, we find that differ by
less than $0.2\%$ on the timescales relevant for our study (\ie $t_{\rm
  FB, N} > 1\,{\rm s}$), thus fully justifying our use of the Newtonian
estimates throughout.

\end{document}